\documentclass[amsmath,amssymb,showpacs,twocolumn,superscriptaddress,pr]{revtex4-1}
\usepackage{graphicx,bm,color,subfigure}
\usepackage[T1]{fontenc}
\usepackage{mathrsfs}
\usepackage{bm}
\usepackage{amsmath}
\usepackage{amssymb}
\usepackage{graphicx}
\usepackage{esint}
\usepackage{multirow}
\usepackage{float}
\usepackage{array}
\usepackage{makecell}
\usepackage{harpoon}
\usepackage{booktabs}
\usepackage{gensymb}
\usepackage{simplewick}
\usepackage{subfigure}


\usepackage[english]{babel}

\begin{document}
\title{High-Angular-Momentum Topological Superconductivity in the Largest-Angle Twisted Homo-bilayer Systems}
\author{Yu-Bo Liu}
\thanks{These two authors contributed equally to this work.}
\affiliation{School of Physics, Beijing Institute of Technology, Beijing 100081, China}
\author{Yongyou Zhang}
\thanks{These two authors contributed equally to this work.}
\affiliation{School of Physics, Beijing Institute of Technology, Beijing 100081, China}
\author{Wei-Qiang Chen}
\affiliation{Shenzhen Key Laboratory of Advanced Quantum Functional Materials and Devices, Southern University of Science and Technology, Shenzhen 518055, China}
\affiliation{Institute for Quantum Science and Engineering and Department of Physics, Southern University of Science and Technology, Shenzhen 518055, China}
\author{Fan Yang}
\email{yangfan_blg@bit.edu.cn}
\affiliation{School of Physics, Beijing Institute of Technology, Beijing 100081, China}
\date{\today}
\begin{abstract}
We study the largest-angle twisted homo-bilayer (LA-THB) systems, hosting Moir\'eless quasi-crystal (QC) structure. We propose to use these materials to generate high-angular-momentum (HAM) topological superconductivities (TSCs) protected by their QC symmetries absent on conventional crystalline materials. This proposal is based on our universal Ginzburg-Landau theory based analysis which yields the general conclusion that, when each $D_n$-symmetric ($n$ is even) monolayer hosts SC with pairing angular momentum $l\le \frac{n}{2}$, the interlayer Josephson coupling will induce SC with pairing angular momentum $L=l$ or $L=n-l$ in the LA-THB, determined by microscopic details. The latter one is just the HAM TSC if $l>0$. Based on our revised perturbational-band theory, we develop general microscopic framework to study the QC LA-THBs involving electron-electron interactions, adopting which we study three examples, i.e. the 30\degree- twisted bilayer graphene, the 30\degree- twisted bilayer BC$_3$, and the 45\degree- twisted bilayer cuprates. The $g+ig$- $h+ih$- and $d+id$- TSCs with HAM $L=4,5$ and $2$ can emerge in certain doping regimes in these systems, respectively.
\end{abstract}

\maketitle
{\bf Introduction:}
``{\it Twistronics}'' is a rapidly emerging new research area\cite{Kennes2021}. In the past several years, there is a surge in the synthesizations and studies of the twisted multi-layer van der Waals heterostructures \cite{Dean2018, caoyuan20181,caoyuan20182, Chenguorui20191,P_Kim2020,Chenshen2020,Park2021,Xian2019,Wang2020,Regan2020,Tang2020}. Motivated by simulating strongly-correlated systems, most of the studies are focused on the small ``magic''-angle twist, which brings about emergent Moir\'e flat bands and reveals various intriguing quantum phases driven by electron correlations\cite{caoyuan20181,caoyuan20182,Chenguorui20191,P_Kim2020,Chenshen2020,Park2021,Xian2019,Wang2020,Regan2020,Tang2020,Yankowitz2019,Yazdani2019,Efetov2019,David2019,Serlin2019,P_Kim2020,Zeldov2020,Caoyuan2021}, arousing tremendous interests \cite{Xu2018,Po2018,Yuan2018,YangFan2018,WuFeng20181,Kang2018,Isobe2018,Koshino2018,Fernandes2018,Dai2019,Gonzalez2019,Song2019,Angeli2019,Linyuping2019, MingXie2020,Abouelkomsan2020,Bultinck2020prx,Senthil2020,Liao2021,Chichinadze2020,Mohammad2020,Xian2021,Angeli2020,Naik2018,Wu_TMD_2018}. Instead, in this paper,we shall present novel quantum states (absent in conventional crystalline materials) in the large angle twist systems. We consider the largest-angle twisted homo-bilayer(LA-THB) system, i.e. two identical $n$-folded rotation-symmetric monolayers stacked with the largest possible twist angle $\frac{\pi}{n}$. Synthesized examples of LA-THBs include the 30\degree-twisted bilayer graphene (TBG) \cite{Ahn2018,Yao2018,Pezzini2020,Yan2019,Deng2020} and the 45\degree-twisted bilayer cuprates\cite{Zhu2021, Zhao2021}. Such materials possess remarkable Moir\'eless quasi-crystal (QC) structures with doubly enlarged rotation symmetries absent on periodic lattices.  It's interesting to investigate the consequence of such enlarged rotation symmetries in the LA-THBs. Though their single-particle properties have been studied\cite{Moon2013,Koshino2015,Moon2019,Park2019,Crosse2021,Yuanshengjun2019,Yuanshengjun20201,Yuanshengjun20202,Aragon2019}, physical properties driven by electron-electron (e-e) interaction have not been studied. Here we propose to use these materials to generate symmetry-protected high-angular-momentum (HAM) topological superconductivities (TSCs), absent in crystalline materials.


The TSCs on 2D lattices are usually generated through the $1:i$ mixing between two degenerate pairing gap functions belonging to a 2D irreducible representations (IRRPs) of the point group, see the Supplementary Material (SM)\cite{SM}. On periodic lattices with at most 6-folded rotation axes, the TSCs could only be the $p+ip$ with pairing angular momentum $L=1$ and the $d+id$ with $L=2$. However, in the QC LA-THB with $2n$-folded rotation symmetry, the HAM TSC with $L\ge3$ is allowed if $n\ge 4$. The SC on QCs has recently been synthesized\cite{exp} and its pairing mechanism has been studied\cite{DeGottardi2013,YuPeng2013,Loring2016,Sakai2017,Andrade2019,Sakai2019,Varjas2019,Nagai2020,Caoye2020,Sakai2020,YYZhang2020,Hauck2021,Zhoubin2021} on intrinsic QCs such as the Penrose lattice. Instead, the LA-THBs belong to extrinsic QCs \cite{Moon2019}, which has a quasi-periodic nature because of the weak coupling between the two crystalline monolayers. In such a system, once each monolayer gets SC through some pairing mechanism, the homo-bilayer can acquire SC through the interlayer Josephson coupling.


In this paper, we first present a study based on the Ginzburg-Landau (G-L) theory, in which we consider a LA-THB formed by two $D_n$-symmetric ($n$ is even here and hereafter) monolayers with each hosting SC with pairing angular momentum $l\le\frac{n}{2}$. The G-L theory yields that the LA-THB would carry SC with pairing angular momentum $L=l$ or $L=n-l\ge\frac{n}{2}$, determined by microscopic details. The latter is just a HAM TSC if $l>0$. Then we develop a microscopic framework suitable to study the QC LA-THBs involving e-e interactions, based on our revised perturbational-band theory. Through microscopic calculations, we predict novel HAM TSCs in three examples, i.e. the QC-TBG, the 30\degree-twisted bilayer BC$_3$ and the 45\degree-twisted bilayer cuprates, which can host $g+ig$- ($L=4$), $h+ih$- ($L=5$), and $d+id$- TSCs in certain doping regimes, respectively.


{\bf Analysis based on G-L theory:} We start from the classification of pairing symmetries on a 2D lattice according to the IRRPs of its $D_{n}$ point group\cite{SM}. It's known that $D_{n}$ has four 1D IRRPs and $\left(\frac{n}{2}-1\right)$ 2D ones (labeled as $E_L$ ($L\in\left[1,\frac{n}{2}-1\right]$)). For each 2D IRRP $E_L$, the two degenerate basis gap functions would generally be mixed as $1:\pm i$ to lower the free energy. The resultant gap function $\Delta^{(\pm)}_L(\mathbf{k})$ transform as $\Delta^{(\pm)}_L(\mathbf{k})\to e^{\mp iL\Delta\phi}\Delta^{(\pm)}_L(\mathbf{k})$ under a $\Delta\phi=2\pi/n$ rotation, corresponding to a TSC with pairing angular momentum $L\le \frac{n}{2}-1$, and pairing chirality ``$+$'' or ``$-$''. The four 1D IRRPs correspond to the non-topological $A_{1,2}$ pairing symmetry with $L=0$ and $B_{1,2}$ one with $L=\frac{n}{2}$.

Then we consider a LA-THB formed by two $D_n$-symmetric monolayers. The symmetry of the LA-THB can be described by the point group $D_{nd}$, which is isomorphic to $D_{2n}$. Assume that driven by some pairing mechanism, the monolayer $\mu=\text{t/b}$ can host a pairing state with ``complex pairing amplitudes'' $\psi_{\mu}$ and normalized gap form factors $\Gamma^{(\mu)}_{l}(\mathbf{k})$, and hence gap function
\begin{equation}\label{gap_function}
\Delta^{(\mu)}(\mathbf{k})=\psi_{\mu}\Gamma^{(\mu)}_{l}(\mathbf{k}).
\end{equation}
Here $l\le \frac{n}{2}$ labels the pairing angular momentum. We shall investigate the pairing symmetry of the LA-THB induced by interlayer Josephson coupling\cite{SM}. The pairing chiralities from both layers are assumed identical\cite{footnote2}, and are set as ``+'' below, without losing generality.

Defining $\hat{P}_{\phi}$ as the rotation by the angle $\phi$, we can set
\begin{equation}\label{relation_gap_updn}
\Gamma^{(\text{b})}_{l}(\mathbf{k})=\hat{P}_{\frac{\pi}{n}}\Gamma^{(\text{t})}_{l}(\mathbf{k}),~~ \hat{P}_{\frac{2\pi}{n}}\Gamma^{(\mu)}_{l}(\mathbf{k})=e^{-i\frac{2l\pi}{n}}\Gamma^{(\mu)}_{l}(\mathbf{k}).
\end{equation}
The symmetry-allowed free energy $F$ as function of $\psi_{\text{t/b}}$ can be decomposed into the monolayers $F_0(\left|\psi_{\mu}\right|^2)$ term and the interlayer Josephson coupling $F_J$ term as\cite{SM}
\begin{eqnarray}\label{G_L_F}
F\left(\psi_{\text{t}},\psi_{\text{b}}\right)&=&F_0(\left|\psi_{\text{t}}\right|^2)+F_0(\left|\psi_{\text{b}}\right|^2)+F_{J}\left(\psi_{\text{t}},\psi_{\text{b}}\right),\nonumber\\F_{J}\left(\psi_{\text{t}},\psi_{\text{b}}\right)&=&-A\left(e^{i\theta}\psi_{\text{t}}\psi_{\text{b}}^*+c.c\right)+O\left(\psi^4\right).
\end{eqnarray}

There is an additional symmetry in the QC LA-THB, i.e. the rotation by $\Delta\phi=\frac{\pi}{n}$ followed by a succeeding layer exchange. Under such combined operations, the gap function on the $\mu$ layer changes from $\Delta^{(\mu)}(\mathbf{k})=\psi_\mu\Gamma_l^{\left(\mu\right)}(\mathbf{k})$ to $\tilde{\Delta}^{(\mu)}(\mathbf{k})=\psi_{\bar{\mu}}\hat{P}_{\frac{\pi}{n}}\Gamma_l^{\left(\bar\mu\right)}(\mathbf{k})$ which, under the relation (\ref{relation_gap_updn}), can be rewritten as $\tilde{\psi}_\mu\Gamma_l^{\left(\mu\right)}(\mathbf{k})$ with
\begin{equation}\label{ODP_change}
\tilde{\psi_\text{b}}=\psi_\text{t}, ~~~~~~~~~~\tilde{\psi_\text{t}}=e^{-i\frac{2l\pi}{n}}\psi_\text{b}.
\end{equation}
This symmetry requires $F(\tilde{\psi_\text{t}},\tilde{\psi_\text{b}})=F(\psi_\text{t},\psi_\text{b})$, dictating
\begin{equation}\label{angle}
e^{i(\theta-\frac{2l\pi}{n})}=e^{-i\theta}\Rightarrow\theta=\frac{l\pi}{n}, ~~~(\text{setting}~ \theta\in\left[0,\pi\right)).
\end{equation}
In this homo-bilayer, the free energy given by (\ref{G_L_F}) should be minimized at $\psi_\text{b}=\pm e^{i\theta}\psi_\text{t}$ for positive/negative $A$ respectively. Consequently, Eq. (\ref{ODP_change}) dictates $(\tilde{\psi_\text{t}}, \tilde{\psi_\text{b}})=\pm e^{-i\theta}(\psi_\text{t},\psi_\text{b})$ and hence $(\tilde{\Delta}^{(\text{t})}(\mathbf{k}),\tilde{\Delta}^{(\text{b})}(\mathbf{k}))=\pm e^{-i\theta}(\Delta^{(\text{t})}(\mathbf{k}),\Delta^{(\text{b})}(\mathbf{k}))$, suggesting that the pairing angular momentum in the LA-THB should be  $L=\theta/\Delta\phi=l$ for $A>0$ or $L=\left|\theta-\pi\right|/\Delta\phi=n-l$ for $A<0$. The latter one corresponds to the HAM TSC if $l>0$.

Note that in the special case of $l=n/2$, we get $\theta=\pi/2$ from  Eq. (\ref{angle}). However, since the monolayer pairing state for $l=n/2$ belongs to 1D $B_{1,2}$ IRRP hosting real and nondegenerate gap form factor, the resultant first-order interlayer Josephson coupling term in Eq. (\ref{G_L_F}) conflicts with the time-reversal symmetry\cite{SM}, and hence should be abandoned. Thus one needs to consider the second-order Josephson coupling term\cite{SM}
\begin{eqnarray}\label{G_L_F_2}
F_J\left(\psi_{\text{t}},\psi_{\text{b}}\right)=-B\left(\psi_{\text{t}}^2\psi_{\text{b}}^{2*}+c.c\right)+O\left(\psi^6\right).
\end{eqnarray}
Eq. (\ref{G_L_F_2}) is minimized at $\psi_b=\pm \psi_t$ for $B>0$ or $\psi_b=\pm i \psi_t$ for $B<0$, with the latter forming a HAM TSC with $L=\frac{n}{2}$ in the LA-THB, belonging to $E_{n/2}$ IRRP of $D_{2n}$.

{\bf Summarizing the G-L theory, for $\bf{l\in \left[1,\frac{n}{2}-1\right]}$ (or $\bf{l=\frac{n}{2}}$), HAM TSC with $\bf{L=n-l> \frac{n}{2}}$ (or $\bf{L=\frac{n}{2}}$) emerges in the LA-THB if the first- (or second-) order interlayer Josephson coupling coefficient $\bf{A<0}$ (or $\bf{B<0}$).} The signs of these coefficients are determined by microscopic calculations.

{\bf Microscopic framework:} The tight-binding (TB) Hamiltonians of our LA-THB take the real-space single-orbital form of $H_{\text{TB}}=\sum_{\mathbf{ij}\sigma}t_{\mathbf{ij}}c^{\dagger}_{\mathbf{i}\sigma}c_{\mathbf{j}\sigma}$, with $t_{\mathbf{ij}}$ provided in the SM\cite{SM}. This Hamiltonian is decomposed into the intra-layer term $H_0$ and the interlayer term $H'$ as
\begin{eqnarray}\label{H_perturb}
H_{0}&=&\sum_{\mathbf{k}\mu\alpha\sigma}c^{\dagger}_{\mathbf{k}\mu\alpha\sigma}c_{\mathbf{k}\mu\alpha\sigma}\varepsilon_{\mathbf{k}}^{\mu\alpha},\nonumber\\
H'&=&\sum_{\mathbf{kq}\alpha\beta\sigma}c^{\dagger}_{\mathbf{k}\text{t}\alpha\sigma}c_{\mathbf{q}\text{b}\beta\sigma}T^{\alpha\beta}_{\mathbf{kq}}+h.c.
\end{eqnarray}
Here $\mathbf{k}/\mathbf{q}$, $\mu(=\text{t(top)},\text{b(bottom)})$, $\alpha/\beta$ and $\sigma$ label the momentum, layer, band and spin respectively, $\varepsilon_{\mathbf{k}}^{\mu\alpha}$ is the monolayer dispersion and $T^{\alpha\beta}_{\mathbf{kq}}$ is given by\cite{MacDonald2011,Castro2007,Moon2013,Koshino2015,Moon2019}
\begin{equation}\label{tunneling_coefficient}
T^{\alpha\beta}_{\mathbf{kq}}=\left\langle \mathbf{k}\alpha^{(\text{t})}\left|H_{\text{TB}}\right|\mathbf{q}\beta^{(\text{b})}\right\rangle.
\end{equation}
In thermal dynamic limit, the nonzero $T^{\alpha\beta}_{\mathbf{kq}}$ requires $\mathbf{k}+\mathbf{G}^{(\text{t})}=\mathbf{q}+\mathbf{G}^{(\text{b})}$\cite{Ahn2018,Yao2018,Moon2013,Koshino2015,Moon2019}, where $\mathbf{G}^{(\text{t/b})}$ represent the reciprocal lattice vectors for the t/b layers. Under this condition, we have
$T^{\alpha\beta}_{\mathbf{kq}}\propto t\left(\mathbf{k}+\mathbf{G}^{(\text{t})}\right)$, which decays promptly with $|\mathbf{k+G}^{(\text{t})}|$. Therefore each zeroth-order top-layer eigenstate $\left|\mathbf{k}\alpha^{(\text{t})}\right\rangle$ can only couple with a few isolated bottom-layer eigenstates $\left|\mathbf{q}\beta^{(\text{b})}\right\rangle$, and vice versa.

\begin{figure}[htbp]
\centering
\includegraphics[width=0.5\textwidth]{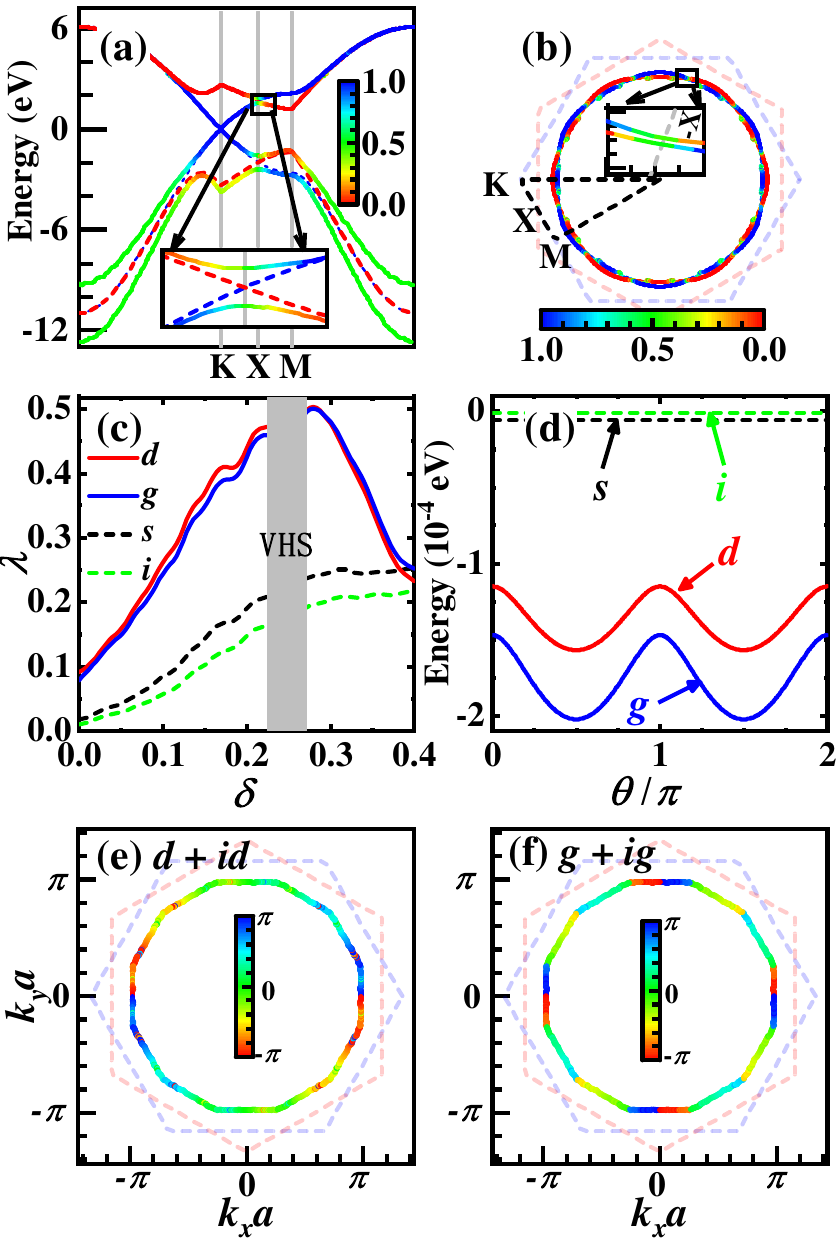}
\caption{(Color on line) (a) Band structure along the high-symmetric lines: solid (dashed) lines for the QC-TBG (two uncoupled graphene monolayers). (b) FSs in the BZ, with the high-symmetric points marked. The colors in (a, b) represent layer component. Insets in (a): band structure near the $X$ point and (b): FSs crossing the $\Gamma$-$X$ line (grey dotted line). (c) Doping $\delta$ dependences of the largest pairing eigenvalues $\lambda$ for the four leading pairing symmetries for $\delta\in (0,0.4)$. The VH doping regime marked grey has been excluded. (d) The angle $\theta$ dependence of the energies for the mixed state from the $d$- and $g$- wave pairings. The energies of the $s$- and $i$- wave pairings are also shown in comparison. The distributions of the gap phases for the $d+id$- (e) and the $g+ig$- (f) TSCs on the inner FS. The doping for (b) and (d) - (f) is $\delta=0.32$.}\label{QC_TBG}
\end{figure}

Note that on a finite lattice, the momenta are discrete, and hence no $\mathbf{q}$ in the bottom layer can satisfy $\mathbf{k}+\mathbf{G}^{(\text{t})}=\mathbf{q}+\mathbf{G}^{(\text{b})}$ for a general $\mathbf{k}$ in the top layer. Therefore for each top-layer state $|\mathbf{k}\alpha^{(\text{t})}\rangle$, we ignore this constraint and directly use Eq. (\ref{tunneling_coefficient}) to find the bottom-layer states $|\mathbf{q}_i\beta_i^{(\text{b})}\rangle$ which obviously couple with it. Then, for these $|\mathbf{q}_i\beta_i^{(\text{b})}\rangle$ states, we find again all the $|\mathbf{k}'_j\alpha_j^{(\text{t})}\rangle$ states which obviously couple with them. Gathering all these states related to $|\mathbf{k}\alpha^{(\text{t})}\rangle$ as bases to form a close sub-space, we can diagonalize the Hamiltonian matrix in this sub-space to obtain all the eigenstates. Among these states, the one having the largest overlap with $|\mathbf{k}\alpha^{(\text{t})}\rangle$ is marked as its perturbation-corrected state $|\widetilde{\mathbf{k}\alpha^{(\text{t})}}\rangle$, whose energy is marked as $\tilde{\varepsilon}^{\text{t}\alpha}_{\mathbf{k}}$. Similarly, we get $|\widetilde{\mathbf{q}\beta^{(\text{b})}}\rangle$ and $\tilde{\varepsilon}^{\text{b}\beta}_{\mathbf{q}}$. This approach can be viewed as a finite-lattice revision of the second-order perturbational-band theory\cite{Yao2018,Koshino2015}.

We have checked that different $|\widetilde{\mathbf{k}\alpha^{(\mu)}}\rangle$ thus obtained are almost mutually orthogonal, qualifying them as a good set of bases for succeeding studies involving e-e interactions. For each system we shall study, the effective pairing interaction obtained in the real space is transformed to the $\{|\widetilde{\mathbf{k}\alpha^{(\mu)}}\rangle\}$ basis and projected onto the Fermi surfaces (FSs). Then we solve the linearized gap equation near the SC $T_c$ to obtain the pairing eigenvalues $\lambda$ and eigenvectors. The pairing symmetry is determined by the eigenvector(s) corresponding to the largest $\lambda$\cite{SM}.

Under this microscopic framework, we study the pairing symmetries in the following three examples. For the LA-THBs of graphene and cuprates with intermediate and strong e-e interactions, we adopt the effective t-J models, treated by the Gutzwiller mean-field approach. For that of BC$_3$, we adopt the small-$U$ Hubbard model treated by the random-phase-approximation (RPA) approach. See the SM\cite{SM} for more information of them.

{\bf Three examples:} The first example is the QC-TBG synthesized recently. Its dodecagonal symmetric QC structure has been verified by various experiments\cite{Ahn2018,Yao2018,Pezzini2020,Yan2019,Deng2020}. Actually, it has been long to search SC in the graphene. Particularly, various groups have predicted \cite{Doniach2007, Gonzalez2008, Honerkamp2008, Pathak2010, McChesney2010, Nandkishore2012, Wang2012, Kiesel2012,Honerkamp2014} the $d+id$ TSC driven by e-e interaction in the monolayer graphene near the VH doping $\delta_{\text{v}}=\frac{1}{4}$ (per unit-cell per spin). Recently, the graphene has been successfully doped to the beyond-VH regime \cite{exp_VHS}, which puts on the agenda the search of the exotic $d+id$ TSC. Based on our G-L theory, the $g+ig$ TSC with $L=6-2=4$ can emerge in the QC-TBG, which however needs to be verified by microscopic calculations.

The band structure of the QC-TBG (solid lines) is shown in Fig. \ref{QC_TBG} (a) along the high-symmetric lines in the Brillouin zone (BZ), in comparison with the uncoupled band structures (dashed lines) from the two separate layers. Remarkably, the band splitting on the electron-doped side is much weaker than that on the hole-doped side, reflecting the much weaker interlayer hybridization on this side\cite{SM}. Our perturbational treatment is focused on the electron-doped side, where the main effect of the interlayer coupling is that when the two uncoupled bands cross at the $X$ points (or in general, the $\Gamma-X$ lines) due to the symmetry, they would hybridize, leading to band splitting and exchange of layer-component. See the inset of Fig. \ref{QC_TBG} (a). Similarly, the two uncoupled sextuple-symmetric Fermi surfaces (FSs) also cross at the $\Gamma-X$ lines and are hybridized into two split dodecagonal-symmetric FSs, with each FS containing equal components from both layers, see Fig. \ref{QC_TBG} (b) and the inset for the electron doping level $\delta=0.32$.

\begin{figure}[htbp]
\centering
\includegraphics[width=0.5\textwidth]{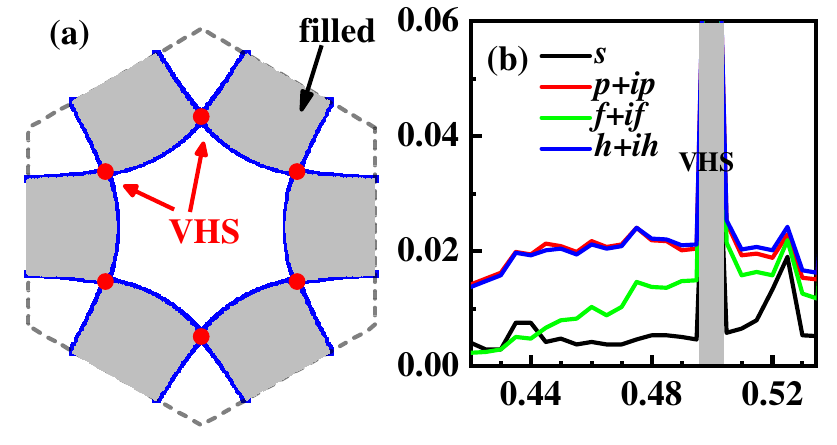}
\caption{(Color on line) (a) The FS of the BC$_3$ doped to its type-II VHS, with the VH momenta $\mathbf{k}_{\text{v}}\ne -\mathbf{k}_{\text{v}}$. The grey-colored regime is filled. (b) Doping $\delta$ dependence of the largest $\lambda$ for the four leading pairing symmetries around the VHS, with the VH doping regime (grey colored) excluded.}\label{BC3_result}
\end{figure}
Figure \ref{QC_TBG}(c) shows the $\delta$ dependence of the largest $\lambda$ of the four leading pairing symmetries in the experimentally accessible doping regime $\delta\in(0,0.4)$, with the VHS regime excluded, as the divergent DOS there might have led to other instabilities. Fig. \ref{QC_TBG}(c) shows that due to the  FS-topology change across the VHS, the leading pairing symmetry changes from the degenerate $d$-wave beneath $\delta_{\text{v}}$ to the degnerate $g$-wave beyond $\delta_{\text{v}}$.  The two components of the $d$- or $g$-wave pairings are mixed as $1:\alpha e^{i\theta}$, and consequently the ground-state energies shown in Fig. \ref{QC_TBG}(d) are minimized at $\alpha=1$ and $\theta=\pm\pi/2$, leading to fully-gapped $d+id$- or $g+ig$- TSCs. The distributions of their gap phases in Figs. \ref{QC_TBG}(e) and (f) on the inner pocket illustrate the winding numbers $2$ and $4$ respectively.

The second example is the 30\degree-twisted bilayer BC$_3$. The BC$_3$ is a graphene-like genuine 2D material already synthesized\cite{BC3_exp}. While the undoped BC$_3$ is a band insulator, it can be electron-doped through chemical absorption with lithium adatoms\cite{BC3_doping}. The low-energy part of the DFT band structure of the electron-doped BC$_3$ is well fitted by a single Boron-$p_z$-orbital TB model on the honeycomb lattice\cite{BC3_VHS}. Remarkably, at the critical doping $\delta_{\text{v}}\sim 1/2$, its FS goes through a Lifshitz transition at which it has six saddle points inside the BZ, as shown in Fig. \ref{BC3_result}(a), forming the type-II VHS\cite{VHS_II}. The combined renormalization-group and RPA approaches predict $p+ip$ TSC near this VH doping via the K-L pairing mechanism\cite{BC3_VHS}. It's interesting to ask: can the $h+ih$ TSC with HAM $L=6-1=5$ be realized in this LA-THB?

The band structure and FSs of this material illustrate similar interlayer hybridization effects as those in the QC-TBG\cite{SM}. The point group and pairing-symmetry classification of this material are identical with those of the QC-TBG. Fig. \ref{BC3_result}(b) exhibits the $\lambda\sim\delta$ relation for the four leading pairing symmetries around $\delta_{\text{v}}$. Clearly, the leading pairing symmetry around $\delta_{\text{v}}$ is just $h+ih$, with its gap-phase winding number to be 5\cite{SM}.

\begin{figure}[htbp]
\centering
\includegraphics[width=0.5\textwidth]{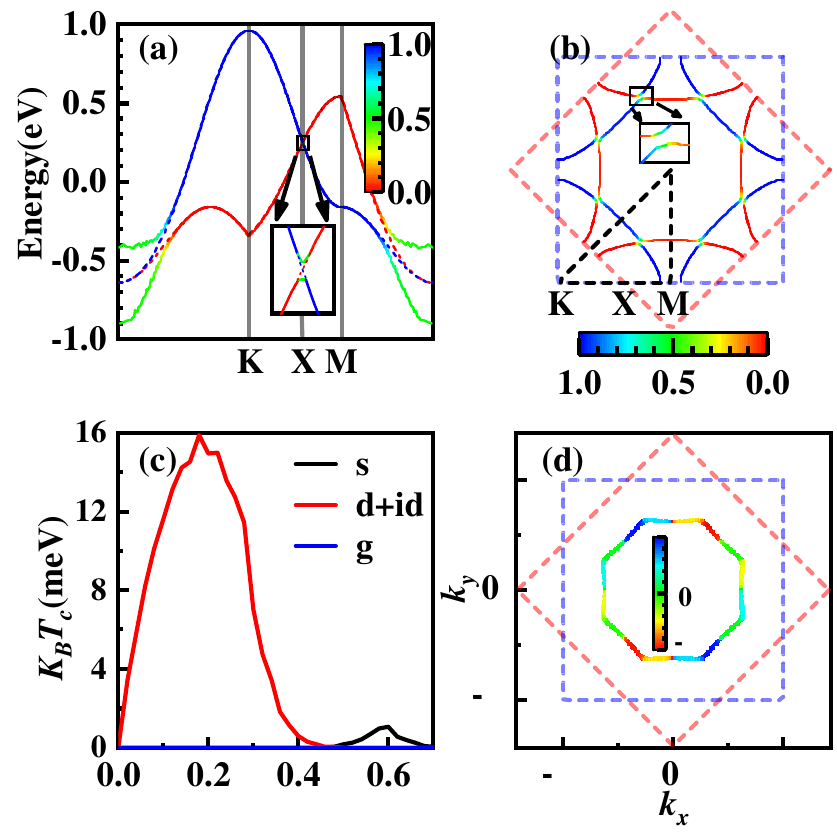}
\caption{(Color on line) Band structure (a) and FSs for $\delta=0.1$ hole-doping (b) of the 45\degree-twisted bilayer cuprates, with the same conventions as introduced in Fig. \ref{QC_TBG} (a) and (b). (c) Doping dependences of the $T_c$ of the three leading pairing symmetries for the hole doping. (d) The distribution of the gap phase of the obtained $d+id$ TSC on the inner pocket.}\label{QC_cuprate}
\end{figure}

The third example is the 45\degree-twisted bilayer cuprates. The DFT calculations predicted the stability of this structure\cite{cuprates_QC}, which was recently experimentally fabricated, and definite evidences for coherent interlayer Josephson tunneling were detected\cite{Zhu2021, Zhao2021}. In the theoretical aspect, although the G-L theory has predicted $d+id$ TSC in this system\cite{cuprates_QC}, more information on the pairing state should be determined by microscopic calculations, which presently are limited to commensurate twisted angles. For general twisted angles, particularly the exact 45\degree, our perturbational-band theory applies.

The band structure and the octagonal-symmetric FSs shown in Fig. \ref{QC_cuprate} (a) and (b) illustrate similar splitting phenomena as those of the QC-TBG caused by interlayer hybridization. The $T_c\sim \delta$ relation for various pairing symmetries belonging to the $D_8$ IRRPs shown in Fig. \ref{QC_cuprate} (c) suggests that the degenerate $(d_{x^2-y^2}, d_{xy})$ doublets are the leading one, which are further mixed as $1:\pm i$ to lower the energy, forming the fully-gapped $d+id$ TSC. The distribution of its gap phase on the inner pocket shown in Fig. \ref{QC_cuprate} (d) exhibits a winding number 2.

{\bf Conclusion and Discussions:} In conclusion, we have provided both universal G-L theory and general microscopic framework to study pairing states driven by interlayer Josephson coupling in the LA-THBs, and obtain the following general conclusions. When each monolayer hosts SC with $l\le \frac{n}{2}$, the LA-THB can carry SCs with $L=l$ or $n-l$, with the latter to be HAM TSC if $l>0$. The first two examples belong to this case with $n=6$ and $l=2,1$, yielding the $g+ig$- and $h+ih$- TSCs with HAM $L=4$ and $L=5$. When each monolayer hosts non-topological SC with $l=\frac{n}{2}$, the LA-THB would carry TSC with $L=\frac{n}{2}$. The last example belongs to this case with $n=4$, yielding the $d+id$ TSC. One more possible example belonging to this case: Ref\cite{f_honeycomb} obtained the parameter regime for the $f$-wave SC on the honeycomb lattice. For this system, our theory predicts the $f+if$ TSC with $L=3$ in the corresponding LA-THB.

Note that on periodic lattices, the HAM TSC can also emerge as the higher-harmonics basis function of the IRRP\cite{Chichinadze2020}. However, in such cases it would generally be considerably mixed with the low-angular-momentum one belonging to the same IRRP, unless the mixing weight of the latter happens to be very small, which is rare. Here our obtained HAM TSC is protected by the QC symmetry not to mix with other pairing states. What's more, the pairing gap function and the topological properties of the obtained TSCs are robust against slight deviation of the twist angle from the largest one\cite{SM}.

In comparison with real-space approaches in the study of SCs on intrinsic QCs\cite{DeGottardi2013,YuPeng2013,Loring2016,Sakai2017,Andrade2019,Sakai2019,Varjas2019,Nagai2020,Caoye2020,Sakai2020,YYZhang2020,Hauck2021,Zhoubin2021}, our $\mathbf{k}$-space perturbational-band theory based microscopic framework lends us more convenience and insight, as we can intuitively examine the distribution of the pairing gap function on the FS. This framework can also be used to study other electron instabilities in the LA-THBs or in other multi-layer heterostructures with arbitrary large twist angles, to find more novel quantum phases.

\section*{Acknowledgements}
We acknowledge stimulating discussions with Cheng-Cheng Liu, Ye Cao, Zheng-Cheng Gu and Yan-Xia Xing. This work is supported by the NSFC under the Grant Nos. 12074031, 12074037, 11674025,11861161001. W.-Q. Chen is supported by the Science, Technology and Innovation Commission of Shenzhen Municipality (No. ZDSYS20190902092905285), Guangdong Basic and Applied Basic Research Foundation under Grant No. 2020B1515120100 and Center for Computational Science and Engineering of Southern University of Science and Technology.

\newpage

\renewcommand{\theequation}{S\arabic{equation}}
\setcounter{equation}{0}
\renewcommand{\thefigure}{S\arabic{figure}}
\setcounter{figure}{0}
\renewcommand{\thetable}{S\arabic{table}}
\setcounter{table}{0}
\begin{widetext}

\begin{table*}[htbp]
	\caption{\label{tab:classification} Irreducible representations (IRRPs) of the point group $D_{n}$ (n is even) and corresponding classification of the pairing symmetries. $D_{n}$ includes four one-dimensional (1D) IRRPs ($A_1, A_2, B_1, B_2$) and $\frac{n}{2}-1$ two-dimensional (2D) IRRPs ($E_1$ to $E_{\frac{n}{2}-1}$). The operator $\hat{P}_{\theta}$ denotes the rotation by the angle $\theta=\frac{2\pi}{n}$ about the $z$- axis and the operator $\hat{\sigma}_y$ denotes the mirror reflection about the $xz$-plane. Here the $x$-axis is set along the direction pointing from the center to the midpoint of an edge of the n-polygon. $D_{(\hat O)}$ is the matrix representation of $D_{n}$ group operators $\hat O$. $C_{\frac{2\pi}{n}}$ and $\sigma_{y}$ are the two generators of $D_{n}$. For each 2D IRRP, the ground-state gap function $\Delta_{\mathbf{k}}=\Delta^{(1)}_{\mathbf{k}}\pm i\Delta^{(2)}_{\mathbf{k}}$, with $\Delta^{(1)}_{\mathbf{k}}$ and $\Delta^{(2)}_{\mathbf{k}}$ being the two basis functions of the 2D IRRP.}
	\begin{center}
	\begin{tabular}{|p{0.03\textwidth}<{\centering}p{0.03\textwidth}<{\centering}|p{0.18\textwidth}<{\centering}|p{0.09\textwidth}<{\centering}|p{0.28\textwidth}<{\centering}|p{0.27\textwidth}<{\centering}|}
		\hline
		\multicolumn{2}{|c|}{IRRPs}         &     $D_{(C_{2\pi/n})}$                                      &   $D_{(\sigma_{y})}$                   &  Basis Functions         & Ground-State Gap Functions\\
		\hline
		\multirow{4}{*}{1D}      & \multicolumn{1}{|c|}{$A_1$} & $I$        &  $I$                      &  $1$
		& $\Delta_{\hat{P}_{\theta}\mathbf{k}}=\Delta_{\mathbf{k}}$, $\Delta_{\hat{\sigma}_{y} \mathbf{k}}=\Delta_{\mathbf{k}}$ \\
		\cline{2-6}
		& \multicolumn{1}{|c|}{$A_2$} & $I$         &  $-I$                  &  {$\left(C^{1}_{n/2}x^{\frac{n}{2}-1}y-C^{3}_{n/2}x^{\frac{n}{2}-3}y^{3}+...\right) * \left(x^{\frac{n}{2}}-C_{n/2}^{2}x^{\frac{n}{2}-2}y^{2}+...\right)$}
		& $\Delta_{\hat{P}_{\theta} \mathbf{k}}=\Delta_{\mathbf{k}}$, $\Delta_{\hat{\sigma}_{y} \mathbf{k}}=-\Delta_{\mathbf{k}}$ \\
		\cline{2-6}
		& \multicolumn{1}{|c|}{$B_1$} & $-I$         &  $I$                  &  ${x^{\frac{n}{2}}-C_{n/2}^{2}x^{\frac{n}{2}-2}y^{2}+...}$
		& $\Delta_{\hat{P}_{\theta} \mathbf{k}}=-\Delta_{\mathbf{k}}$, $\Delta_{\hat{\sigma}_{y} \mathbf{k}}=\Delta_{\mathbf{k}}$ \\
		\cline{2-6}
		& \multicolumn{1}{|c|}{$B_2$} & $-I$         &  $-I$                  &  ${C^{1}_{n/2}x^{\frac{n}{2}-1}y-C^{3}_{n/2}x^{\frac{n}{2}-3}y^{3}+...}$
		& $\Delta_{\hat{P}_{\theta} \mathbf{k}}=-\Delta_{\mathbf{k}}$, $\Delta_{\hat{\sigma}_{y} \mathbf{k}}=-\Delta_{\mathbf{k}}$ \\
		\hline
		\multirow{5}{*}{2D}      & \multicolumn{1}{|c|}{$E_1$} & $I\cos\frac{2\pi}{n} - i\sigma_{y}\sin\frac{2\pi}{n}$                      &  $\sigma_{z}$ & $(x, y)$
		& $\Delta_{\hat{P}_{\theta}\mathbf{k}}=e^{\pm i{2\pi\over n}}\Delta_{\mathbf{k}}$, $\Delta_{\hat{\sigma}_{y} \mathbf{k}} =\Delta^{*}_{\mathbf{k}}$ \\
		\cline{2-6}
		& \multicolumn{1}{|c|}{$...$} &  $...$    &  $...$               &  $...$
		&$...$ \\
		\cline{2-6}
			& \multicolumn{1}{|c|}{$E_L$} &  $I\cos\frac{2L\pi}{n} - i\sigma_{y}\sin\frac{2L\pi}{n}$    &  $\sigma_{z}$               &  $(x^{L}-C^{2}_{L}x^{L-2}y^{2}+... ,\ C_{L}^{1}x^{L-1}y-C^{3}_{L}x^{L-3}y^{3}+...)$
		&$\Delta_{\hat{P}_{\theta}\mathbf{k}}=e^{\pm i{2L\pi\over n}}\Delta_{\mathbf{k}}$, $\Delta_{\hat{\sigma}_{y} \mathbf{k}} =\Delta^{*}_{\mathbf{k}}$ \\
		\cline{2-6}
		& \multicolumn{1}{|c|}{$...$} &  $...$    &  $...$               &  $...$
		&$...$ \\
		\cline{2-6}
		& \multicolumn{1}{|c|}{$E_{\frac{n}{2}-1}$} &  $I\cos\frac{(n-2)\pi}{n} - i\sigma_{y}\sin\frac{(n-2)\pi}{n}$    &  $\sigma_{z}$               &  $(x^{\frac{n}{2}-1}-C^{2}_{n/2-1}x^{\frac{n}{2}-3}y^{2}+... ,   C_{n/2-1}^{1}x^{\frac{n}{2}-2}y-C^{3}_{n/2-1}x^{\frac{n}{2}-4}y^{3}+...)$
		&$\Delta_{\hat{P}_{\theta}\mathbf{k}}=e^{\pm i{(n-2)\pi\over n}}\Delta_{\mathbf{k}}$, $\Delta_{\hat{\sigma}_{y} \mathbf{k}} =\Delta^{*}_{\mathbf{k}}$ \\
		\hline
	\end{tabular}
	\end{center}
	\label{d12group}
\end{table*}

\section{Pairing-symmetry classification on the $D_{n}$-symmetric lattices}

This section introduces the classification of the pairing symmetries on the $D_{n}$-symmetric ($n$ is even here and hereafter) lattices according to the irreducible representations (IRRPs) of the point group $D_{n}$, see Table \ref{d12group}~\cite{x3}.

In Table \ref{d12group}, the second and third columns list the representation matrices of the two generators of $D_{n}$ up to a global unitary transformation. The fourth column provides the basis functions of each IRRP. For the 1D IRRPs, the basis function of identity representation $A_1$ is 1, the basis functions of $B_1$ and $B_2$ are the real and imaginary parts of $(x+iy)^{\frac{n}{2}}$, respectively. Their product gives the basis function of $A_2$. The two basis functions of the 2D IRRP $E_L$ are the real and imaginary parts of $(x+iy)^L$. In the last column of Table \ref{d12group}, we show the properties of the ground-state pairing gap function of each pairing symmetry. For each 1D IRRP, the normalized ground-state gap functions is taken as the basis function. For each 2D IRRP, the normalized ground-state gap function is the $1:\pm i$ mixing of the two basis functions of that IRRP. Such a mixing manner can be understood from the following Ginzburg-Landau (G-L) theory in combination with our numerical calculations.

Setting $\psi_1$ and $\psi_2$ as the two global ``complex amplitudes'' in front of the two normalized gap form factors taken as the two basis functions in the considered 2D IRRP, the G-L free-energy function takes the following form,
\begin{eqnarray}
F(\psi_1, \psi_2) = C_1\left(|\psi_{1}|^2+|\psi_{2}|^2\right)+
C_2\left|\psi_{1}^{2}+\psi_{2}^{2}\right|^{2}+
C_3\left(\left|\psi_{1}\right|^{2}+\left|\psi_{2}\right|^{2}\right)^2 + O(|\psi|^{6}).
\end{eqnarray}
 Notice that this formula is consistent with that obtained in Ref.~\cite{x2}, which satisfies all the symmetries in the system, including the time-reversal, the U(1)-gauge and the point-group symmetries. We neglect the $O(|\psi|^{6})$ term and assume the minimized free energy is realized at $\psi_{2}=e^{i\theta}\psi_{1}$, then
\begin{eqnarray}
F(\psi_1, \psi_2)  = 2C_1|\psi_1|^2 + 2C_2|\psi_{1}|^{4} [\cos (2\theta)+1] + 4C_3|\psi_1|^4.
\end{eqnarray}
When $C_2>0$ the minimization of $F$ requires $\theta=\pm{\pi\over 2}$, that is, $\psi_{1}:\psi_{2}=1: (\pm i)$. This is why our calculation results can well be fitted with the cosine function (see later in Fig.~\ref{fit}). When $C_2<0$ the minimization of $F$ requires $\theta=0$ or $\pi$, that is, $\psi_{1}:\psi_{2}=1: (\pm 1)$. Although from the G-L theory alone, one doesn't know the sign of $C_2$, physically the solution $\psi_{1}:\psi_{2}=1: (\pm i)$ is more reasonable because in such cases the obtained pairing gap function is fully gapped, which benefits the energy gain.

For the 2D IRRP $E_L$, the $1:\pm i$ mixings between the two basis functions lead to complex ground-state gap functions, whose complex phase angles change $\pm L\theta$ with each rotation by the angle $\theta=2\pi/n$. These states each forms a 1D IRRP of the $C_{n}$ subgroup, with the quantity $L$ just to be the pairing angular momentum, and the sign ``$\pm$'' to be the pairing chirality. In the meantime, the mirror reflection symmetry $\sigma$ is broken in these mixed states. The pairing angular momentum of the 1D $A_{1,2}$ and $B_{1,2}$ IRRPs can also be defined as the ratio of the change of the pairing-gap phase over the rotation angle, which consequently leads to $L=0$ and $L=\frac{n}{2}$ for the former and latter IRRPs, respectively.

The $D_{n}$ point groups relevant to the three examples studied in our work, i.e. the 30\degree-twisted bilayer graphene, the 30\degree-twisted bilayer BC$_3$, the 45\degree-twisted bilayer cuprates, include the $D_6$,$D_4$,$D_{12}$ and $D_{8}$. Firstly, there are $D_6$ symmetric monolayers(the first two examples) and $D_4$ symmetric monolayers(the last example) in our work. For $D_6$, the 1D IRRPs include the $A_1$ (the $s$-wave with $L=0$), the  $A_2$ ($f{*}f'$-wave, $L=0$), the $B_1, B_2$ ($f$- and $f'$-, $L=\frac{n}{2}=3$); and the 2D IRRPs include the $E_L\ (L=1,2)$ (the $p$,$d$), respectively. For $D_{4}$, the 1D IRRPs include the $A_1$ ($s$-wave, $L=0$), $A_2$ ($d{*}d'$-, $L=0$), $B_1, B_2$ ($d$- and $d'$-, $L=\frac{n}{2}=2$); and the 2D IRRPs include the $E_L\ (L=1)$ (the $p$) respectively. Secondly, after the two monolayers are stacked with the largest possible twist angle, the resultant largest-angle twisted homo-bilayer (LA-HTB) systems are $D_{6d}$ (isomorphic to $D_{12}$) symmetric and $D_{4d}$ (isomorphic to $D_8$) symmetric. For $D_{12}$, the 1D IRRPs include the $A_1$ (the $s$-wave with $L=0$), the  $A_2$ ($i{*}i'$-wave, $L=0$), the $B_1, B_2$ ($i$- and $i'$-, $L=\frac{n}{2}=6$); and the 2D IRRPs include the $E_L\ (L=1,2, \cdots, 5)$ (the $p$, $d$, $f$, $g$, $h$), respectively. For $D_{8}$, the 1D IRRPs include the $A_1$ ($s$-wave, $L=0$), $A_2$ ($g{*}g'$-, $L=0$), $B_1, B_2$ ($g$- and $g'$-, $L=\frac{n}{2}=4$); and the 2D IRRPs include the $E_L\ (L=1,2, 3)$ (the $p$, $d$, $f$) respectively.

\section{More details on the Ginzburg-Landau theory}

This section provides some details omitted in the main text for the G-L theory with interlayer Josephson coupling, mainly including the formula of the G-L free-energy functions and the proof that they satisfy all the symmetries of the system. Here we consider two $D_n$-symmetric monolayers coupled into a LA-THB, with the monolayer $\mu=\text{t(top)/b(bottom)}$ hosting a pairing state with ``complex pairing amplitudes'' $\psi_{\mu}$ and normalized gap form factors $\Gamma^{(\mu)}_{l}(\mathbf{k})$, and hence gap function
\begin{equation}\label{gap_function}
\Delta^{(\mu)}(\mathbf{k})=\psi_{\mu}\Gamma^{(\mu)}_{l}(\mathbf{k}).
\end{equation}
Here $l\le \frac{n}{2}$ labels the pairing angular momentum. We shall investigate the pairing symmetry of the LA-THB induced by interlayer Josephson coupling. The following analysis will be divided into two parts, with one part for $n/2-1\ge l \ge 1 $ in which the monolayer pairing states are doubly degenerate, and the other for $l=0, n/2$ in which the monolayer pairing state is non-degenerate.

\subsection{The degenerate cases of $n/2-1\ge l \ge 1 $}
This subsection considers the cases of $n/2-1\ge l \ge 1 $, in which each layer in the LA-THB hosts two degenerate pairings belonging to the 2D $E_l$ IRRPs. In such cases, we first let the two degenerate pairing gap functions within each layer to be mixed as $1:\pm i$ to form chiral pairing state belonging to $E_{l}^{1}\pm iE_{l}^{2}$ IRRP, and then consider the interlayer Josephson coupling. The reason for such consideration lies in that the interlayer coupling in the LA-THB is weak and can be treated as perturbation. Further more, we only consider the cases wherein the pairings chiralities from the two monolayers are identical, i.e. both layers take $E_{l}^{1}+ iE_{l}^{2}$ or $E_{l}^{1}- iE_{l}^{2}$, because otherwise the system cannot gain energy from the interlayer Josephson coupling. The argument for this point is as follow.

Physically, the interlayer Josephson coupling originates from the second-order perturbational process with taking the interlayer tunneling term $H'$ in Eq.~(\ref{perturbation}) as perturbation. The contribution of this second-order perturbation to the free energy can be approximated as
\begin{eqnarray}\label{Josephson0}
F_{interlayer}\approx -\frac{\left\langle H'^2\right\rangle}{2\Delta},\qquad
H'= -\sum_{\textbf{i}\textbf{j}\sigma}  c_{\textbf{i}\rm{t}\sigma}^{\dagger} c_{\textbf{j}\rm{b}\sigma}t_{\textbf{i}\textbf{j}}+h.c.,
\end{eqnarray}
where $\Delta$ denotes the averaged pairing-gap amplitude. The Eq. (\ref{Josephson0}) can be Wick-decomposed as
\begin{eqnarray}\label{Josephson1}
F_{interlayer}&\approx&
 -\frac{1}{2\Delta}\sum_{\mathbf{i}\mathbf{j}\tilde{\mathbf{i}}\tilde{\mathbf{j}}} \Delta_{\mathbf{i}\tilde{\mathbf{i}}}^{(\rm t)*} \Delta_{\mathbf{j}\tilde{\mathbf{j}}}^{(\rm b)} t_{\textbf{i}\textbf{j}} t_{\tilde{\textbf{i}}\tilde{\textbf{j}}}+c.c.
\end{eqnarray}
 Now if the pairing chiralities for $\Delta^{(\text{b})}_{\mathbf{j}\tilde{\mathbf{j}}}$ and $\Delta^{(\text{t})}_{\mathbf{i}\tilde{\mathbf{i}}}$ are opposite, we can, without lossing generality, let $\Delta^{(\text{b})}_{\mathbf{j}\tilde{\mathbf{j}}}\sim \Delta_{E_{l}^{1}+ iE_{l}^{2}}$ and $\Delta^{(\text{t})}_{\mathbf{i}\tilde{\mathbf{i}}}\sim \Delta_{E_{l}^{1}- iE_{l}^{2}}$. By symmetry, we have
\begin{eqnarray}\label{symmetry}
\Delta^{(\text{b})}_{\hat P_{\frac{2\pi}{n}}\mathbf{j}\hat P_{\frac{2\pi}{n}}\tilde{\mathbf{j}}}=e^{i\frac{2l\pi}{n}}\Delta^{(\text{b})}_{\mathbf{j}\tilde{\mathbf{j}}},~~~~\Delta^{(\text{t})}_{\hat P_{\frac{2\pi}{n}}\mathbf{i}\hat P_{\frac{2\pi}{n}}\tilde{\mathbf{i}}}=e^{-i\frac{2l\pi}{n}}\Delta^{(\text{t})}_{\mathbf{i}\tilde{\mathbf{i}}},~~~~
t_{\hat P_{\frac{2\pi}{n}}\mathbf{i}\hat P_{\frac{2\pi}{n}}\mathbf{j}}=t_{\mathbf{i}\mathbf{j}},t_{\hat P_{\frac{2\pi}{n}}\tilde{\mathbf{i}}\hat P_{\frac{2\pi}{n}}\tilde{\mathbf{j}}}=t_{\tilde{\mathbf{i}}\tilde{\mathbf{j}}}.
\end{eqnarray}
Then from Eq. (\ref{Josephson1}), we have
\begin{eqnarray}\label{noncoupling}
F_{interlayer} \approx -\frac{1}{2\Delta}\sum_{m=1}^n e^{i\frac{4l\pi}{n}m}{\sum_{\mathbf{i}}}'\sum_{\mathbf{j}\tilde{\mathbf{i}}\tilde{\mathbf{j}}} \Delta_{\mathbf{i}\tilde{\mathbf{i}}}^{(\rm t)*} \Delta_{\mathbf{j}\tilde{\mathbf{j}}}^{(\rm b)} t_{\textbf{i}\textbf{j}} t_{\tilde{\textbf{i}}\tilde{\textbf{j}}}+c.c. =0 +c.c.=0,
\end{eqnarray}
where ${\sum_{\mathbf{i}}}'$ represents the sum of $\frac{1}{n}$ of all the sites. Eq. (\ref{noncoupling}) suggests that if the pairing chiralities of the two monolayers are opposite, the system could not gain energy from the interlayer Josephson coupling. Therefore, we only consider the case wherein the pairings in the two monolayers both belong to the $E_{l}^{1}+ iE_{l}^{2}$ or $E_{l}^{1}- iE_{l}^{2}$ IRRP.

When the pairing states on both monolayers belong to the $E_{l}^{1}\pm iE_{l}^{2}$ IRRP, we can let their form factors satisfy
\begin{eqnarray}
	\Gamma_{l,E_{l}^{1}\pm iE_{l}^{2}}^{(\rm b)}=\hat{P}_{\frac{\pi}{n}}\Gamma_{l,E_{l}^{1}\pm iE_{l}^{2}}^{(\rm t)},~~ \hat{P}_{\frac{2\pi}{n}}\Gamma_{l,E_{l}^{1}\pm iE_{l}^{2}}^{(\rm \mu)}=e^{\mp i\frac{2l\pi}{n}}\Gamma_{l,E_{l}^{1}\pm iE_{l}^{2}}^{(\rm \mu)}.	
\end{eqnarray}
Here $\hat{P}_\phi$ denotes the rotation by the angle $\phi$. The free energy $F$ as function of the complex amplitudes $\psi_{\text{t/b}}$ reads,
\begin{eqnarray}\label{fEdplusid}
F_{E_{l}^{1}\pm iE_{l}^{2}}(\psi_{\rm t},\psi_{\rm b})=F_{0} (|\psi_{\rm t}|^2)+F_{0}(|\psi_{\rm b}|^2)-A(e^{\pm i\theta}\psi_{\rm t}\psi_{\rm b}^*+c.c)+O(\psi^4).
\end{eqnarray}
Here the $F_0$ and the $A$ terms denote the contributions from each monolayer and their Josephson coupling, respectively. Here we have set $A\in R$ and $\theta \in [0,\pi)$. The above form of $F_{E_{l}^{1}+ iE_{l}^{2}}$ and $F_{E_{l}^{1}- iE_{l}^{2}}$ are time-reversal related.

In the following, we verify that Eq. (\ref{fEdplusid})  satisfies  all symmetries of the system, including the time-reversal, the U(1)-gauge, and the point-group symmetries. The time-reversal operation dictates
\begin{eqnarray}
	\psi_{(\rm t/\rm b)} \rightarrow \tilde\psi_{(\rm t/\rm b)}=	\psi_{(\rm t/\rm b)}^*, \qquad
	\Gamma_{l,E_{l}^{1}+ iE_{l}^{2}}^{(\rm t/\rm b)} \rightarrow \Gamma_{l,E_{l}^{1}+ iE_{l}^{2}}^{(\rm t/\rm b)*}=\Gamma_{l,E_{l}^{1}- iE_{l}^{2}}^{(\rm t/\rm b)}.
\end{eqnarray}
The U(1)-gauge transformation dictates
\begin{eqnarray}\label{u1}
	\psi_{(\rm t/\rm b)} \rightarrow \tilde\psi_{(\rm t/\rm b)}=	e^{i\eta}\psi_{(\rm t/\rm b)},
\end{eqnarray}
where $\eta$ is an arbitrary phase angle. For the point-group operations, we only need to consider the two generators of the group: one is the rotation by the angle $\phi=\frac{\pi}{n}$, followed by a succeeding layer exchange, and the other can be chosen as the specular reflection operation $\sigma$ that changes the layer index. The former generator dictates
\begin{eqnarray}\label{ry}
\psi_{\rm t} \rightarrow \tilde\psi_{\rm t}=	e^{-i\frac{2l\pi}{n}}\psi_{\rm b},\qquad
\psi_{\rm b} \rightarrow \tilde\psi_{\rm b}=	\psi_{\rm t}.
\end{eqnarray}
The latter generator dictates
\begin{eqnarray}
	\psi_{\rm t} \rightarrow \tilde\psi_{\rm t}=	\psi_{\rm b},
	\qquad
	\psi_{\rm b} \rightarrow \tilde\psi_{\rm b}=	\psi_{\rm t},
	\qquad
	\Gamma_{l,E_{l}^{1}+ iE_{l}^{2}}^{(\rm t/\rm b)} \rightarrow \Gamma_{l,E_{l}^{1}- iE_{l}^{2}}^{(\rm t/\rm b)}.
\end{eqnarray}
It can be checked that the free-energy function (\ref{fEdplusid}) satisfies all these symmetries under the condition $\theta=\frac{l\pi}{n}$.

The remaining part of the G-L analysis has been provided in the main text. Briefly, $F$ is minimized at $\psi_\text{b}=e^{il\pi/n}\psi_\text{t}$ for $A>0$ or $\psi_\text{b}=-e^{il\pi/n}\psi_\text{t}$ for $A<0$, leading to a bilayer pairing state belonging to the $E_L$ IRRP of $D_{2n}$ with distinguished pairing angular momentum $L=l$ for $A>0$ or $L=n-l$ for $A<0$. The latter one is the HAM TSC.

\subsection{The non-degenerate cases of $l=0, n/2$}
This subsection considers the cases of $l=0$ and $n/2$, in which the pairing in each monolayer belong to 1D $A_{1,2}$ and $B_{1,2}$ IRRP, respectively. In such cases, the pairing gap form factor on each monolayer is real and non-degenerate.

Let's first expand the free-energy function to the first-order interlayer Josephson coupling terms as,
\begin{eqnarray}\label{f_second}
F_{A/B}(\psi_{\rm t},\psi_{\rm b})=F_{0} (|\psi_{\rm t}|^2)+F_{0}(|\psi_{\rm b}|^2)-A(e^{i\theta}\psi_{\rm t}\psi_{\rm b}^*+c.c)+O(\psi^4).
\end{eqnarray}
This function should satisfy all the symmetries of the system. The U(1)-gauge symmetry has been satisfied. Since the monolayer gap form factor is real, the invariance under the time-reversal operation $\psi_{(\rm t/\rm b)} \rightarrow \psi_{(\rm t/\rm b)}^*$ dictates $e^{i\theta}=\pm 1$. The $\frac{\pi}{n}$ rotation (followed by a layer exchange) described in Eq. (\ref{ry}) causes $\psi_{\rm t} \rightarrow \tilde\psi_{\rm t}=\psi_{\rm b},\psi_{\rm b} \rightarrow \tilde\psi_{\rm b}=	\psi_{\rm t}$ for $l=0$ (the $A_{1,2}$ IRRP) and $\psi_{\rm t} \rightarrow \tilde\psi_{\rm t}=	-\psi_{\rm b},\psi_{\rm b} \rightarrow \tilde\psi_{\rm b}=	\psi_{\rm t}$ for $l=n/2$ (the $B_{1,2}$ IRRP). For $l=0$, Eq. (\ref{f_second}) is invariant under such operation and the remaining specular reflection operation, which therefore satisfies all the symmetries of the system. However, for $l=n/2$, the invariance of Eq. (\ref{f_second}) under such operation requires $A=0$, which suggests that we should expand the free-energy function to the second-order interlayer Josephson coupling.

For $l=n/2$, the free-energy function is expanded to the second-order interlayer Josephson coupling as
\begin{eqnarray}\label{fEdplusid2}
F_{B}(\psi_{\rm t},\psi_{\rm b})=F_{0} (|\psi_{\rm t}|^2)+F_{0}(|\psi_{\rm b}|^2)-B\left(\psi_{\text{t}}^2\psi_{\text{b}}^{2*}+c.c\right)+O\left(\psi^6\right).
\end{eqnarray}
This formula satisfies the U(1)-gauge symmetry. The time-reversal symmetry requires $B$ to be real number. The $\frac{\pi}{n}$ rotation (followed by a layer exchange) symmetry mentioned above is also easily checked to be satisfied. So does the specular reflection symmetry. Therefore, Eq. (\ref{fEdplusid2}) satisfies all the symmetries of the system for the case of $l=n/2$.

The remaining part of the G-L analysis has been includes in the main text. Briefly, for $l=0$, $F$ is minimized at $\psi_\text{b}=\psi_\text{t}$ for $A>0$ or $\psi_\text{b}=-\psi_\text{t}$ for $A<0$, leading to a bilayer pairing state with distinguished pairing angular momentum $L=0$ for $A>0$ or $L=n$ for $A<0$. The result of $l=0$ is similar with those for $n/2-1\ge l \ge 1 $ obtained on the above. However, for $l=n/2$, $F$ is minimized at $\psi_b=\pm \psi_t$ for $B>0$ or $\psi_b=\pm i \psi_t$ for $B<0$, with the latter forming a bilayer TSC belonging to the 2D $E_{n/2}$ IRRP with pairing angular momentum $L=n/2$. The latter case is supported by microscopic calculations.

\section{perturbational band theory based microscopic framework}
\subsection{Perturbational band theory}
This subsection provides some details of the perturbational-band-theory approach adopted in our study of the LA-THB. The detailed information of the band structures and the Fermi surfaces (FSs) thus obtained for the three exemplar systems are provided in the Sec. IV.

We start from the following tight-binding (TB) model,
\begin{equation}\label{H_TB}
H_{\text{TB}}=-\sum_{\mathbf{ij}\sigma}t_{\mathbf{ij}}c^{\dagger}_{\mathbf{i}\sigma}c_{\mathbf{j}\sigma},
\end{equation}
where $\sigma$ labels spin and $t_{\mathbf{ij}}$ represents the hopping integral between the site $\mathbf{i}$ and site $\mathbf{j}$ which can locate either within intralayer or at interlayer. While the intralayer hopping integrals for the three exemplar systems will be provided separately in the next section, the interlayer ones are unifiedly given as \cite{1}
\begin{equation}\label{tij}
t_{\mathbf{ij}}=t_{\mathbf{ij}\pi}\left [1-\left(\frac{\mathbf{R}_{\mathbf{ij}}\cdot\mathbf{e}_{\mathbf{z}}}{R}\right)^{2}\right]+
t_{\mathbf{ij}\sigma}\left(\frac{\mathbf{R}_{\mathbf{ij}}\cdot\mathbf{e}_{\mathbf{z}}}{R}\right)^{2},
\end{equation}
with
\begin{equation}
t_{\mathbf{ij}\pi}=t_{\pi}e^{-\left(R_{\mathbf{ij}}-a\right)/r_{0}},\quad
t_{\mathbf{ij}\sigma}=t_{\sigma}e^{-(R_{\mathbf{ij}}-d)/r_{0}}.\nonumber
\end{equation}
Here, $R_\mathbf{ij}$ is the length of the 3D vector $\mathbf{R}_\mathbf{ij}$, pointing from site $\mathbf{i}$ to site $\mathbf{j}$ ($\mathbf{i}$ and $\mathbf{j}$ locate at different layers), and $\mathbf{e}_{\mathbf{z}}$ is the unit vector perpendicular to the layer. The parameters $a$, $d$, $r_{0}$, $t_\pi$,  $t_\sigma$ denote the lattice constant, interlayer spacing, normalization distance, in-plane hoping and vertical hoping, respectively.

The Hamiltonian \eqref{H_TB} could be decomposed into the zeroth-order intralayer hopping term $H_0$ and perturbational interlayer tunneling term $H'$, namely,
\begin{eqnarray}\label{perturbation}
H_0=\sum_{\textbf{k}\mu\alpha\sigma}  c_{\textbf{k}\mu\alpha\sigma}^\dag c_{\textbf{k}\mu\alpha\sigma}\varepsilon_{\textbf{k}}^{\mu\alpha},\qquad
H'=\sum_{\textbf{kq}\alpha\beta\sigma}  c_{\textbf{k}{\rm t}\alpha\sigma}^\dag c_{\textbf{q}{\rm b}\beta\sigma}T_{\textbf{kq}}^{\alpha\beta}+h.c.,
\end{eqnarray}
where $\mu\ [=\rm \text{t (top)},\text{b (bottom)}]$ and $\alpha$ are the layer and band indices, respectively. $\textbf{k}$ and $\textbf{q}$ label the momentum, and $\varepsilon_{\textbf{k}}^{\mu\alpha}$ is the dispersion of the single layer. The eigenstate of $H_0$ is denoted as $| \textbf{k}\alpha^{(\mu)}\rangle$ (here we omit the spin index for simplicity, since the following treatment is independent of the spin degree of freedom), representing a monolayer state on the layer $\mu$. The interlayer tunneling matrix element $T_{\textbf{kq}}^{\alpha\beta}$ reads
\begin{eqnarray}    \label{interlayer_hopping}
T_{\textbf{kq}}^{\alpha\beta}=\langle \textbf{k}\alpha^{(\rm t)}|H_{\text{TB}} |\textbf{q}\beta^{(\rm b)}\rangle
= -\frac{1}{N}\sum_{\textbf{ij}}
\xi_{\textbf{i},\textbf{k}{\rm t}\alpha}^{*}
\xi_{\textbf{j},\textbf{q}{\rm b}\beta}
t_{\mathbf{ij}},
\end{eqnarray}
where $\xi/\sqrt{N}$ ($N$ is the number of unit cells on each monolayer.) represents the real-space wave function for the monolayer state $| \textbf{k}\alpha^{(\mu)}\rangle$.

Our approach is a finite-lattice version of the second-order perturbational-band theory\cite{2,3}. Given a zeroth-order state $|\mathbf{k}\alpha^{(\mu)}\rangle$  with the zeroth-order energy $\varepsilon_{\textbf{k}}^{\mu\alpha}$, we provide in the following our procedure to obtain its perturbation-corrected state $|\widetilde{\mathbf{k}\alpha^{(\mu)}}\rangle$, and the perturbation-corrected energy $\tilde{\varepsilon}^{\mu\alpha}_{\mathbf{k}}$. For a state $| \textbf{k}\alpha^{(\rm t)}\rangle$ in the top layer, one can find some states $|\textbf{q}\beta^{(\rm b)}\rangle$ in the bottom layer which couple with it through Eq. (\ref{interlayer_hopping}). In thermal dynamic limit, the nonzero coupling matrix element $T^{\alpha\beta}_{\mathbf{kq}}$ requires~\cite{1,2,3}
\begin{equation}\label{couple_condition}
\mathbf{k}+\mathbf{G}^{(\text{t})}=\mathbf{q}+\mathbf{G}^{(\text{b})},
\end{equation}
where $\mathbf{G}^{(\text{t/b})}$ represent the reciprocal lattice vectors of the top/bottom layers. However, on our finite lattice with discrete momentum points, for each typical $\mathbf{k}$ on the top layer, no $\mathbf{q}$ on the bottom layer can make the relation (\ref{couple_condition}) exactly satisfied unless $\mathbf{k}=\mathbf{q}=\mathbf{G}^{(\text{t})}=\mathbf{G}^{(\text{b})}=0$, as the two layers are mutually incommensurate with each other. Therefore for each $|\mathbf{k}\alpha^{(\text{t})}\rangle$ state on the top layer, we ignore this constraint and directly use Eq. (\ref{interlayer_hopping}) to numerically find the $|\mathbf{q}_i\beta_i^{(\text{b})}\rangle$ states on the bottom layer which obviously couple with it. Here we only keep those $|\mathbf{q}_i\beta_i^{(\text{b})}\rangle$ states when their tunneling strengths $T_{\textbf{kq}}^{\alpha\beta}$ with $|\mathbf{k}\alpha^{(\text{t})}\rangle$ are larger than $0.2$ times of the maximum one in all $T_{\textbf{kq}}^{\alpha\beta}$. One can imagine that the momenta of these kept states only make the relation (\ref{couple_condition}) approximately satisfied. Then, for these $|\mathbf{q}_i\beta_i^{(\text{b})}\rangle$ states, we find again all the $|\mathbf{k}'_j\alpha_j^{(\text{t})}\rangle$ states on the top layer which obviously couple with them. Gathering all these states related to $|\mathbf{k}\alpha^{(\text{t})}\rangle$ as the bases to form a close sub-space, we can write down and diagonalize the Hamiltonian matrix to obtain all the eigenstates. Among these states, the one having the largest overlap with $|\mathbf{k}\alpha^{(\text{t})}\rangle$ is marked as its perturbation-corrected state $|\widetilde{\mathbf{k}\alpha^{(\text{t})}}\rangle$, whose energy is marked as $\tilde{\varepsilon}^{\text{t}\alpha}_{\mathbf{k}}$. The procedure to get $|\widetilde{\mathbf{q}\beta^{(\text{b})}}\rangle$ and $\tilde{\varepsilon}^{\text{b}\beta}_{\mathbf{q}}$ is similar.

Our numerical results verify that different perturbation-corrected states $|\widetilde{\mathbf{k}\alpha^{(\rm\mu)}}\rangle$ are almost orthogonal to each other, which justifies the set of wave functions $\left\{|\widetilde{\mathbf{k}\alpha^{(\rm\mu)}}\rangle\right\}$ as a good basis for the following study involving electron-electron (e-e) interactions.  Writing the creation (annihilation) operator of the eigenstate $|\widetilde{\mathbf{k}\alpha^{(\rm\mu)}}\rangle$ for the spin $\sigma$ as $\tilde{c}^{\dagger}_{\mathbf{k\mu}\alpha\sigma}$ ($\tilde{c}_{\mathbf{k\mu}\alpha\sigma}$), we have
\begin{equation}\label{transform}
\tilde{c}^{\dagger}_{\mathbf{k\mu}\alpha\sigma}=\frac{1}{\sqrt{N}}\sum_{\mathbf{i}}c^{\dagger}_{\mathbf{i}\sigma}\tilde\xi_{\mathbf{i},\mathbf{k\mu}\alpha}, ~~~~~~~~~~c_{\mathbf{i}\sigma}=\frac{1}{\sqrt{N}}\sum_{\mathbf{k\mu}\alpha}\tilde{c}_{\mathbf{k\mu}\alpha\sigma}\tilde\xi_{\mathbf{i},\mathbf{k\mu}\alpha}.
\end{equation}
Here $\tilde\xi_{\mathbf{i},\mathbf{k\mu}\alpha}/\sqrt{N}$ represents the real-space wave function of the perturbation-corrected eigenstate $|\widetilde{\mathbf{k}\alpha^{(\rm\mu)}}\rangle$. The TB model (\ref{H_TB}) can be transformed to this eigen basis as
\begin{equation}
H_{\text{TB}}=\sum_{\mathbf{k}\mu\alpha\sigma}\tilde{\varepsilon}_{\mathbf{k}}^{\mu\alpha}\tilde{c}_{\mathbf{k\mu}\alpha\sigma}^{\dagger}\tilde{c}_{\mathbf{k\mu}\alpha\sigma}.
\end{equation}
The real-space interaction Hamiltonians studied in the following can also be transformed to this eigen-basis.

\subsection{Gutzwiller-MF for the t-J model}

This subsection provides some details of the Gutzwiller mean-field (MF) \cite{Gutzwiller} calculations for the t-J model in the QC-TBG and the 45\degree-twisted bilayer cuprates.

The Hamiltonian of the t-J model reads
\begin{eqnarray}\label{tJ}
H &=& H_{\text{TB}}+H_{J} = -\sum_{\mathbf{ij}\sigma}t_{\mathbf{ij}}c^{\dagger}_{\mathbf{i}\sigma}c_{\mathbf{j}\sigma}+\sum_{\mathbf{i,j}}J_{\mathbf{ij}}\mathbf{S}_{\mathbf{i}}\cdot\mathbf{S}_{\mathbf{j}},
\end{eqnarray}
with $J_{\mathbf{ij}}=4t^2_{\mathbf{ij}}/U$. The parameters $U$ for the two systems will be presented in the Section IV. Here the no-double-occupance constraint is imposed on the Hilbert space.

In the Gutzwiller-MF treatment\cite{Gutzwiller}, the no-double-occupance constraint can be realized as setting $t_{\mathbf{ij}}\to \delta t_{\mathbf{ij}}$, where $\delta$ denotes the doping level deviating from half-filling. Consequently, we get
\begin{eqnarray}\label{Gutzwiller}
H_{\text{G-TB}} = \delta H_{\text{TB}} = -\delta\sum_{\mathbf{ij}\sigma}t_{\mathbf{ij}}c^{\dagger}_{\mathbf{i}\sigma}c_{\mathbf{j}\sigma}=
\delta\sum_{\mathbf{k}\mu\alpha\sigma}\tilde{\varepsilon}_{\mathbf{k}}^{\mu\alpha}\tilde{c}_{\mathbf{k\mu}\alpha\sigma}^{\dagger}\tilde{c}_{\mathbf{k\mu}\alpha\sigma}.
\end{eqnarray}
When we use Eq. (\ref{Gutzwiller}) as the effective kinetic part of the Hamiltonian, we no longer need to consider the no-double-occupance constraint under the Gutzwiller-MF treatment\cite{Gutzwiller}. Then we come to the e-e interaction. As the AFM superexchange interaction in the t-J model favors spin-singlet pairings and suppress spin-triplet pairings, we can write this part of the Hamiltonian as the pre-MF-decomposed formula in the spin-singlet channel as,
\begin{eqnarray}\label{MF}
H_J &=& \sum_{(\mathbf{i,j})}J_{\mathbf{ij}}\mathbf{S}_{\mathbf{i}}\cdot\mathbf{S}_{\mathbf{j}} \to -\frac{3}{4}\sum_{(\mathbf{i,j})}J_{\mathbf{ij}}\Delta_{\mathbf{ij}(0,0)}^{\dagger} \Delta_{\mathbf{ij}(0,0)}, ~~~~~~~~~~\Delta_{\mathbf{ij}(0,0)}=\frac{1}{\sqrt{2}}\left( c_{\mathbf{i}\uparrow} c_{\mathbf{j}\downarrow}- c_{\mathbf{i}\downarrow} c_{\mathbf{j}\uparrow}\right)
\end{eqnarray}
This interaction Hamiltonian can be transformed to the $\left\{|\widetilde{\mathbf{k}\alpha^{(\rm\mu)}}\rangle\right\}$ basis to get the following BCS Hamiltonian,
\begin{eqnarray}    \label{pairingpotentialtJ}
H_J^{(s)} &=&\sum_{\mathbf{k}\mu\alpha\atop\mathbf{q}\nu\beta}
\frac{1}{N}\tilde{c}_{\mathbf{k\mu\alpha}\downarrow}^{\dagger}
\tilde{c}_{\mathbf{-k\mu\alpha}\uparrow}^{\dagger}
\tilde{c}_{\mathbf{-q\nu\beta}\uparrow}
\tilde{c}_{\mathbf{q\nu\beta}\downarrow}
V^{\mu\nu}_{\alpha\beta}(\mathbf{k},\mathbf{q}), \nonumber\\ V^{\mu\nu}_{\alpha\beta}(\mathbf{k},\mathbf{q})&=&\frac{-3}{2N}\sum_{(\mathbf{i,j})}J_{\mathbf{ij}}\mathbf{Re}(\tilde\xi_{\mathbf{i},\mathbf{k}\mu\alpha}\tilde\xi_{\mathbf{j},\mathbf{k}\mu\alpha}^{*})\mathbf{Re}(\tilde\xi_{\mathbf{i},\mathbf{q}\nu\beta}\tilde\xi_{\mathbf{j},\mathbf{q}\nu\beta}^{*}).
\end{eqnarray}
Note that we only consider the intra-band pairing with opposite momenta and spin here, i.e. the pairing between the time-reversal pair $|\widetilde{\mathbf{k}\alpha^{(\rm\mu)}}\uparrow\rangle$ and $|\widetilde{-\mathbf{k}\alpha^{(\rm\mu)}}\downarrow\rangle$. Finally, the effective Gutzwiller-BCS Hamiltonian we need to treat is,
\begin{equation}\label{G-BCS}
H_{\text{G-BCS}}=H_{\text{G-TB}}+H_J^{(s)}.
\end{equation}

The MF treatment on Eq. (\ref{G-BCS}) yields the following linearized gap equation near the superconducting $T_c$\cite{4, 5},
\begin{align}\label{linear_eq}
-\frac{1}{(2\pi)^2}\sum_{\nu\beta}\oint dq_{\parallel}
\frac{V^{\mu\nu}_{\alpha\beta}(\mathbf{k},\mathbf{q})}
{ v^{\nu\beta}_F(\bm{q})}\Delta_{\nu\beta}(\mathbf{q})
=\delta\lambda_{\delta}\Delta_{\mu\alpha}(\mathbf{k})\equiv\lambda\Delta_{\mu\alpha}(\mathbf{k}),
\end{align}
where $v_F^{\nu\beta}(\mathbf{q})$ is the bare Fermi velocity (without imposing the no-double-occupance constraint) and $q_{\parallel}$ denotes the component along the tangent of the FS. For each doping level $\delta$, the pairing symmetry is determined by the normalized gap function $\Delta_{\mu\alpha}(\mathbf{k})$ corresponding to the largest pairing eigenvalue $\lambda$ solved for this equation. The MF pairing temperature $T^*$ is related to $\lambda_\delta$ via the relation $T^*\sim e^{-1/\lambda_\delta}$. Note that the MF pairing temperature $T^*$ is not the real superconducting $T_c$. The quantity $T^*$ only reflects the MF pairing gap amplitude $\Delta_{\text{MF}}$, which is related to the true SC order parameter $\Delta_{\text{SC}}$ via the relation $\Delta_{\text{SC}}\approx\delta\Delta_{\text{MF}}$ in the Gutzwiller-MF treatment\cite{Gutzwiller}. As the SC order parameter scales with the superconducting $T_c$, we have $T_c\approx \delta T^*\propto \delta e^{-1/\lambda_\delta}\propto \delta e^{-\delta/\lambda}$.

\subsection{RPA for the Hubbard model}
This subsection provides the detail of the RPA treatment of the Hubbard model on the 30\degree - twisted BC$_3$ studied in our work. The Hamiltonian of the Hubbard model reads
\begin{eqnarray}\label{Hubbard}
H &=& H_{\text{TB}}+H_{U} = \sum_{\mathbf{k}\mu\alpha\sigma}\tilde{\varepsilon}_{\mathbf{k}}^{\mu\alpha}\tilde{c}_{\mathbf{k\mu}\alpha\sigma}^{\dagger}\tilde{c}_{\mathbf{k\mu}\alpha\sigma}+U\sum_{\mathbf{i}}n_{\mathbf{i}\uparrow}n_{\mathbf{i}\downarrow},
\end{eqnarray}
where $n_{\mathbf{i}\sigma}=c_{\mathbf{i}\sigma}^{\dagger}c_{\mathbf{i}\sigma}$ is the particle number operator and $U>0$ means a repulsive interaction.

To treat this model, the bare susceptibility $\chi^{(0)}$ is defined as
\begin{eqnarray}\label{chi0}
\chi^{(0)}_{\mathbf{i,j}} &=&\int^{\beta}_{0}d\tau e^{i\omega_n\tau}\left\langle T_{\tau}c_{\mathbf{i}}^{\dagger}(\tau)c_{\mathbf{i}}(\tau)c_{\mathbf{j}}^{\dagger}(0) c_{\mathbf{j}}(0)\right\rangle  = \frac{1}{N^2} \sum_{\mathbf{k,q}\atop\mu\nu\alpha\beta}\tilde{\xi}_{\mathbf{i},\mathbf{k\mu}\alpha}^{*}\tilde{\xi}_{\mathbf{j},\mathbf{k\mu}\alpha}\tilde{\xi}_{\mathbf{i},\mathbf{q\nu}\beta}\tilde{\xi}_{\mathbf{j},\mathbf{q\nu}\beta}^{*}\dfrac{n_f(\tilde{\varepsilon}_{\mathbf{k}}^{\mu\alpha}-\mu_c)-n_f(\tilde{\varepsilon}_{\mathbf{q}}^{\nu\beta}-\mu_c)}{\tilde{\varepsilon}_{\mathbf{q}}^{\nu\beta}-\tilde{\varepsilon}_{\mathbf{k}}^{\mu\alpha}}.
\end{eqnarray}
The renormalized $\chi$ in the RPA level reads,
\begin{eqnarray}\label{chi}
\chi =\left(I-U\chi^{(0)}\right)^{-1}\chi^{(0)}.
\end{eqnarray}
Here we have taken $\chi$ and $\chi^{(0)}$ as matrices, whose elements in the $\mathbf{i}$-th row and $\mathbf{j}$-th column are just $\chi_{\mathbf{i,j}}$ or $\chi^{(0)}_{\mathbf{i,j}}$. Then, the effective Hamiltonian of the system via the real-space Kohn-Luttinger mechanism\cite{x1} can be written as
\begin{eqnarray}\label{Heff}
H_{\text{RPA-eff}}=\sum_{\mathbf{k}\mu\alpha\sigma}\tilde{\varepsilon}_{\mathbf{k}}^{\mu\alpha}\tilde{c}_{\mathbf{k\mu}\alpha\sigma}^{\dagger}\tilde{c}_{\mathbf{k\mu}\alpha\sigma}+U\sum_{\mathbf{i}}c_{\mathbf{i}\uparrow}^{\dagger}c_{\mathbf{i}\uparrow}c_{\mathbf{i}\downarrow}^{\dagger}c_{\mathbf{i}\downarrow}-\dfrac{U^2}{2}\sum_{\mathbf{i,j}\atop\sigma\sigma'}c_{\mathbf{i}\sigma}^{\dagger}c_{\mathbf{i}\sigma'}c_{\mathbf{j}\sigma'}^{\dagger}c_{\mathbf{j}\sigma}\chi_{\mathbf{ij}}.
\end{eqnarray}

The following MF processing of Eq.~\eqref{Heff} is parallel to the subsection B. Concretely, we shall first transform this real-space Hamiltonian into the $\mathbf{k}$-space in the $\left\{|\widetilde{\mathbf{k}\alpha^{(\rm\mu)}}\rangle\right\}$ basis. Then through a MF study, we obtain the linearized gap equation at $T_c$ similar with Eq. (\ref{linear_eq}), with only the $V^{\mu\nu}_{\alpha\beta}(\mathbf{k},\mathbf{q})$ in that equation replaced by
\begin{eqnarray}\label{V_hubbard_s}
V^{(s)\mu\nu}_{\alpha\beta}(\mathbf{k},\mathbf{q})=\frac{U}{N}\sum_{\mathbf{i}}|\tilde\xi_{\mathbf{i},\mathbf{k}\mu\alpha}\tilde\xi_{\mathbf{i},\mathbf{q}\nu\beta}|^2+\frac{U^2}{N}\sum_{(\mathbf{i,j})}\chi_{\mathbf{ij}}\mathbf{Re}(\tilde\xi_{\mathbf{i},\mathbf{k}\mu\alpha}\tilde\xi_{\mathbf{j},\mathbf{k}\mu\alpha}^{*})\mathbf{Re}(\tilde\xi_{\mathbf{i},\mathbf{q}\nu\beta}\tilde\xi_{\mathbf{j},\mathbf{q}\nu\beta}^{*}).
\end{eqnarray}
for the singlet pairing and
\begin{eqnarray}\label{V_hubbard_t}
V^{(t)\mu\nu}_{\alpha\beta}(\mathbf{k},\mathbf{q})=-\frac{U^2}{N}\sum_{(\mathbf{i,j})}\chi_{\mathbf{ij}}\mathbf{Im}(\tilde\xi_{\mathbf{i},\mathbf{k}\mu\alpha}\tilde\xi_{\mathbf{j},\mathbf{k}\mu\alpha}^{*})\mathbf{Im}(\tilde\xi_{\mathbf{i},\mathbf{q}\nu\beta}\tilde\xi_{\mathbf{j},\mathbf{q}\nu\beta}^{*}).
\end{eqnarray}
for the triplet one. The leading pairing symmetry is determined by the pairing eigenvector corresponding to the largest pairing eigenvalue $\lambda$, which is related to the $T_c$ via $T_c\propto e^{-1/\lambda}$.

\section{More informations on the numerical results for the three examples}
This section provides more information about the results for the three exemplar systems studied in our work, including the band structures, the FSs, and the obtained gap functions.

\subsection{The QC-TBG}\label{example-I}
\begin{figure}[htbp]
	\centering
	\includegraphics[width=0.95\textwidth]{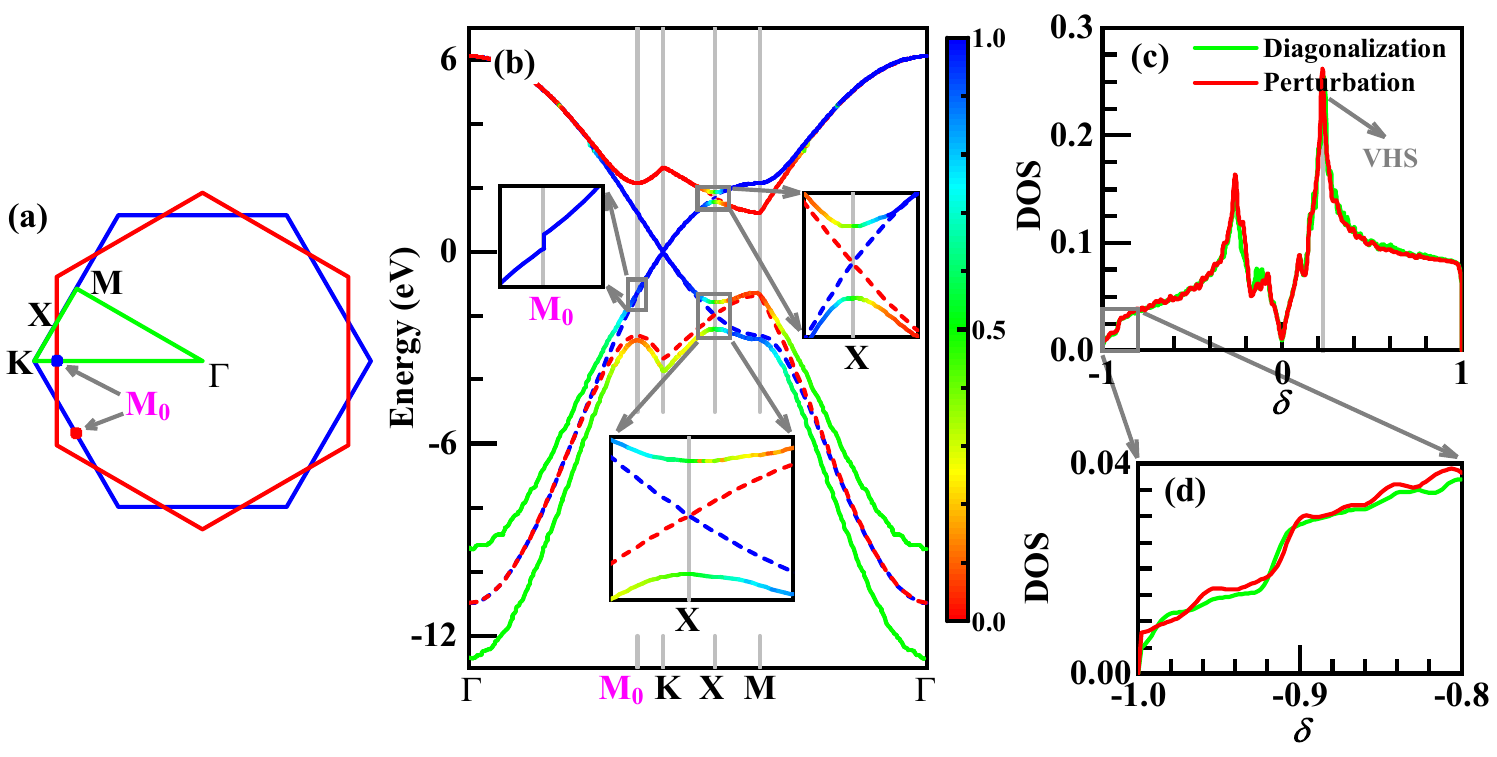}
	\caption{(a) High-symmetry points marked in the Brillouin zone. (b) Band structures of the QC-TBG with interlayer tunneling (solid lines) and without interlayer tunneling (dashed lines) along the high-symmetry lines in (a). (c) DOS calculated by the perturbational-band theory in comparison with that obtained by real-space diagonalization on a finite lattice with 90000 sites. (d) Enlarged view of the DOS around $\delta=-0.9$, presenting the twice step-like drops of the DOS.}
	\label{dos-gap}
\end{figure}

For the QC-TBG, both the intralayer and interlayer hopping integrals are provided by
Eq.~\eqref{tij}, with the related parameters given by $a\approx0.142$ nm, $d\approx0.335$ nm, $t_{\pi}\approx 2.7$ eV, $t_\sigma\approx -0.48$ eV and $r_{0}\approx0.0453$ nm. These band-structure parameters are taken from Ref.~\cite{1}. The interaction parameter $U=10$ eV is taken from Ref\cite{graphene_RMP}

Along the lines connecting the high-symmetry points in the Brillouin zone marked in Fig.~\ref{dos-gap}(a), we plot our obtained band structure (solid lines) in Fig.~\ref{dos-gap}(b), in comparison with the uncoupled band structures (dashed lines) from the two layers. A remarkable feature of  Fig.~\ref{dos-gap}(b) is the obvious particle-hole (p-h) asymmetry: while the band structure on the electron-doped side is overall not far from simply overlaying the two sets of uncoupled monolayer band structures, there is strong interlayer hybridization and split within the band-bottom regime near the $\Gamma$-point on the hole-doped side. This split is reflected by the twice step-like drops in the density of states at the lower filling region (with doping level $\delta\approx-0.9$) in the hole-doped side, as shown in Figs.~\ref{dos-gap}(c) and \ref{dos-gap}(d), which is different from the monolayer graphene and is consistent with the result from  Ref.~\cite{3}. Such a p-h asymmetry is caused by the relatively weaker interlayer coupling on the electron-doped side, as was revealed in Refs.~\cite{1,3} and proved in the following.

Near the $\Gamma$ point, the band structure on each layer comprises two types: one band labeled as ``$+$'' originates from the bonding between the A and B sublattices, and the other labeled as ``$-$'' originates from the anti-bonding between the two sublattices. These four zeroth-order states near the $\Gamma$ point for the two layers read,
\begin{eqnarray}\label{couplings}
|\textbf{k}\pm^{(\text{t/b})}\rangle&\approx&\frac{1}{\sqrt{2}}\left(|\textbf{k}A^{(\text{t/b})}\rangle\pm|\textbf{k}B^{(\text{t/b})}\rangle\right),
\end{eqnarray}
On each layer, the energy of the bonding state (labeled by ``$+$'') is lower than that of the anti-bonding state (labeled by ``$-$''). Therefore, the bonding and anti-bonding states occupy the bottom and top regimes of the band, respectively. Then we consider the interlayer couplings between each two zeroth-order states from different layers.  Before that, we first evaluate the coupling between the states $|\textbf{k}X^{{(\rm t)}}\rangle$ and $|\textbf{q}Y^{(\rm b)}\rangle$~\cite{1} [here $X$ and $ Y (=\rm A,\ B)$ are sublattice indices],
 \begin{eqnarray}\label{matrix_element_AB}
 \langle \textbf{k}X^{{(\rm t)}}|H_{\text{TB}} |\textbf{q}Y^{(\rm b)}\rangle\approx-t(\mathbf{k})\delta_{\mathbf{k},\mathbf{q}},
 \end{eqnarray}
if $\mathbf{k}$ and $\mathbf{q}$ are near the $\Gamma$ point. From Eq.~\eqref{couplings} and Eq.~\eqref{matrix_element_AB}, one gets
\begin{eqnarray}
&\langle\textbf{k}+^{{(\rm t)}}|H_{\text{TB}} |\textbf{q}+^{(\rm b)}\rangle \approx -2t(\mathbf{k})\delta_{\mathbf{k},\mathbf{q}}, \qquad
&\langle\textbf{k}-^{(\rm t)}|H_{\rm TB}|\textbf{q}-^{(\rm b)}\rangle \approx 0.
\end{eqnarray}
Therefore, the strong interlayer coupling only takes place in the bottom regime of the band, which is responsible for the p-h asymmetry character of the band structure. Hereafter, we shall focus on the electron-doped side, because the weak interlayer coupling there validates our perturbational approach.

\begin{figure}[htbp]
	\centering
	\includegraphics[width=0.9\textwidth]{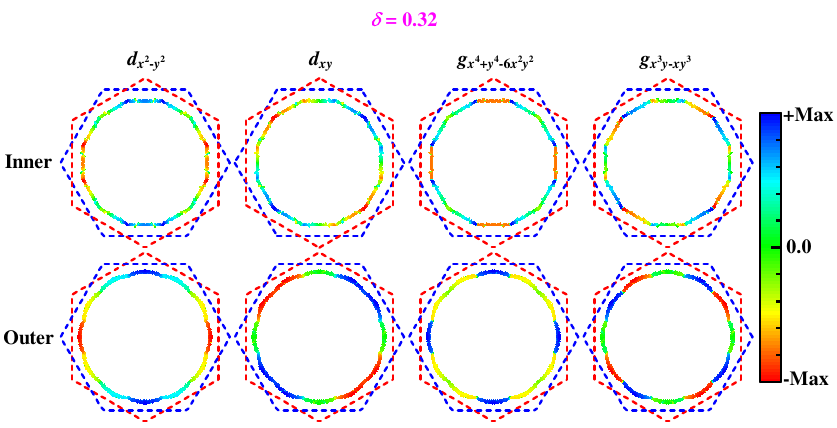}
	\caption{Distributions of pairing gap functions for $d$- and $g$-waves on the inner and outer Fermi surfaces for $\delta=0.32$ electron doping in the QC-TBG.}
	\label{dg}
\end{figure}

The main effect of the interlayer coupling on the electron-doped side lies in that the top-layer band branches and the bottom-layer ones cross and split at the $X$ point. Before and after the $X$ point, the related two bands exchange their layer components, as shown in the inset of Fig.~\ref{dos-gap}(b). Actually, such band crossing and splitting take place on the whole $\Gamma$-$X$ line: for each $\mathbf{k}$ on this line, by symmetry, the states $|\mathbf{k}\alpha^{(\text{t})}\rangle$ and $|\mathbf{k}\alpha^{(\text{b})}\rangle$ possess degenerate zeroth-order energy. They are further coupled via $\mathbf{k}+\mathbf{G}^{(\text{t})}=\mathbf{k}+\mathbf{G}^{(\text{b})}$ by setting $\mathbf{G}^{(\text{t})}=\mathbf{G}^{(\text{b})}=0$. The perturbational coupling between the two degenerate states $|\mathbf{k}\alpha^{(\text{t})}\rangle$ and $|\mathbf{k}\alpha^{(\text{b})}\rangle$ leads to their hybridization with a band split between each other. Consequently, the FSs contributed from the two layers also cross and split once they cross the $\Gamma$-$X$ line. As a consequence of this interlayer coupling, the emergent bonding and anti-bonding FSs possess dodecagonal symmetry. In addition, the tiny gaps (about 0.1eV) at the points $\mathbf{M}_{0}^{(\rm t/b)}=\mathbf{G}^{(\rm b/t)}/2$ are revealed by the insets in Fig.~\ref{dos-gap}(b). These gaps are caused by the second-order perturbational coupling between the states $\left|\mathbf{M}_{0}^{(\mu)}\alpha^{(\mu)}\right\rangle$ and $\left|-\mathbf{M}_{0}^{(\mu)}\alpha^{(\mu)}\right\rangle$, consistent with Ref.~\cite{2}.

Figure~\ref{dg} shows the leading gap functions of the degenerate $(d_{x^2-y^2},d_{xy})$- and $(g_{x^4+y^4-6x^2y^2},g_{x^3y-xy^3})$- wave pairing symmetries corresponding to their largest pairing eigenvalues under $\delta=0.32$ . As the two pockets are close in the Brillioun zone, we plot the distributions of the gap functions on the inner and outer pockets separately to enhance the visibility. Figure~\ref{dg} informs us the following characters of these gap functions. Firstly, while the $d_{x^2-y^2}$- and the $g_{x^4+y^4-6x^2y^2}$- wave pairing gap functions are even with respect to the $x$- and $y$- axes, the $d_{xy}$- and the $g_{x^3y-xy^3}$- wave ones are odd about these axes. Secondly, while the two $d$-wave pairing gap functions change sign for every 90\degree rotation, the two $g$-wave ones keep unchanged for such rotation. Thirdly, while each $d$-wave pairing gap function possesses four nodal points on each pocket, each $g$-wave pairing gap function possesses eight nodal points on each pocket. Finally, the nodal points for the two $d$-wave gap functions don't coincide with each other, and neither do those for the two $g$- wave ones.
\begin{figure}[htbp]
	\centering
	\includegraphics[width=0.5\textwidth]{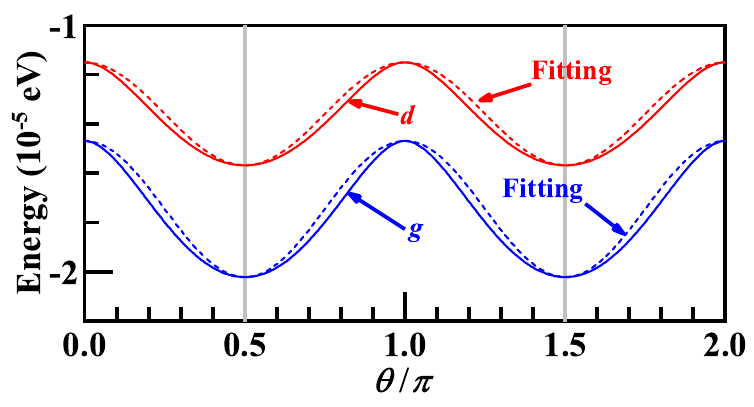}
	\caption{Variations of the energies $E$ with the mixing-phase-angle $\theta$ for the degenerate $d$- and $g$-wave pairings obtained for $\delta=0.32$ electron doping in the QC-TBG, with their global pairing amplitudes optimized for the energy minimization. The dashed lines represent the fittings of the cosine functions with the formula $E\left(\theta\right)=E_0+\eta\cos 2\theta$, with $\eta>0$.}
	\label{fit}
\end{figure}

\begin{figure}[htbp]
	\centering
	\includegraphics[width=0.9\textwidth]{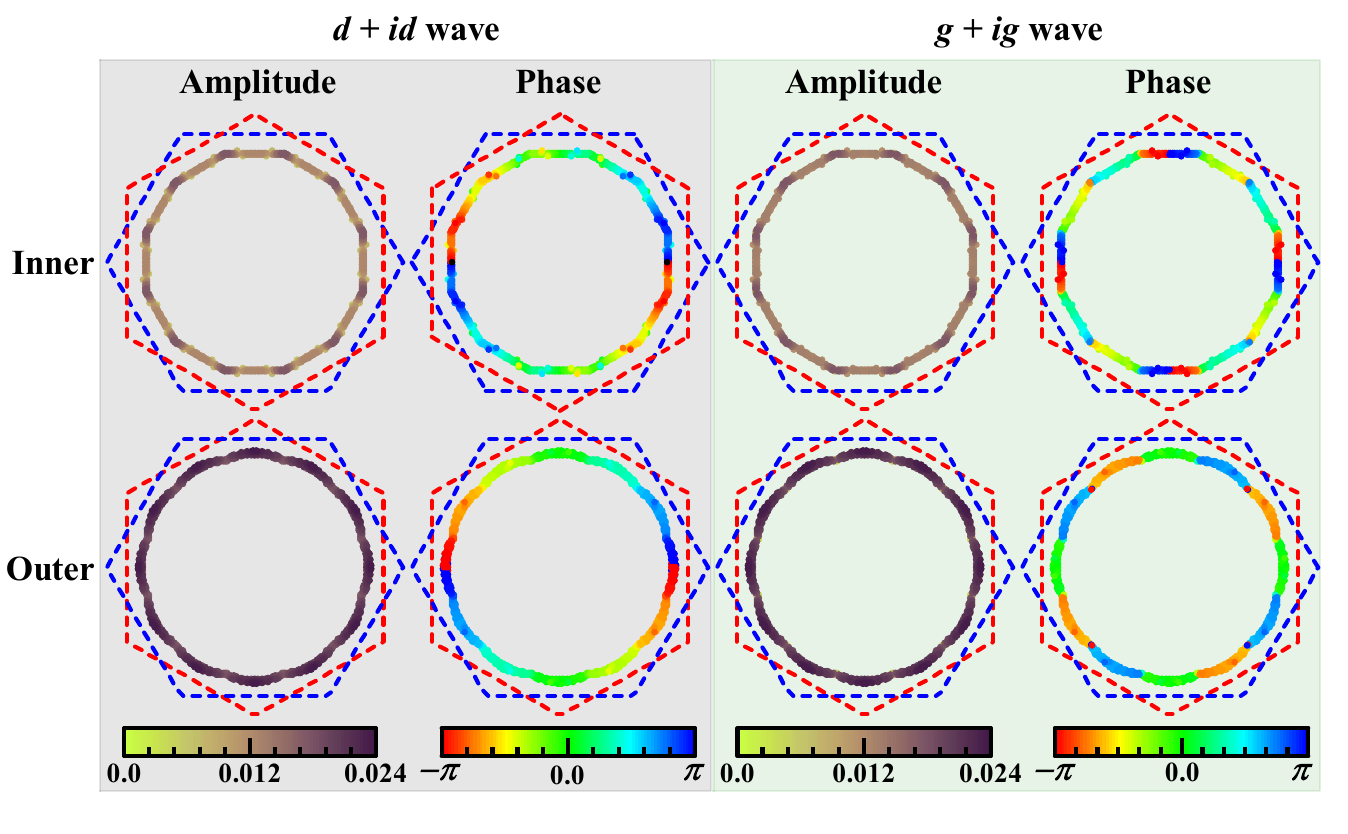}
	\caption{Distributions of the amplitudes and phases of pairing gap functions for the $d+id$- and $g+ig$-waves on the inner and outer Fermi surfaces for $\delta=0.32$ electron doping in the QC-TBG.}
	\label{mixing}
\end{figure}

Since the two $d$- and $g$- wave pairing gap functions each are doubly degenerate, we shall mix the two basis functions for each case to minimize the ground-state energy.  For this purpose,
we write the pairing gap function as $\Delta_{\mu\alpha}(\mathbf{k})=\psi_{1}\Delta^{(1)}_{\mu\alpha}(\mathbf{k})
	+\psi_{2}\Delta^{(2)}_{\mu\alpha}(\mathbf{k})$, where $\psi_{1}$ and $\psi_{2}$ are two complex numbers. The $\Delta^{(1)}_{\mu\alpha}(\mathbf{k})$ and $\Delta^{(2)}_{\mu\alpha}(\mathbf{k})$ represent the two degenerate normalized basis functions of the $d$- wave or $g$- wave pairing symmetries obtained from solving the linearized gap equation (\ref{linear_eq}). Using this pairing gap function, we obtain the BCS-MF Hamiltonian, diagonalizing which we obtain the BCS-MF ground state. Then $\psi_{1}$ and $\psi_{2}$ are determined by minimizing the expectation value $E$ of the Gutzwiller-BCS effective Hamiltonian (\ref{G-BCS}). Setting $\psi_1:\psi_2=1:\alpha e^{i\theta}$, our results suggest $\alpha=1$ and $E$ as a function of $\theta$, i.e. $E(\theta)$ are shown in Fig. \ref{fit} for both the $d$- and $g$- wave pairings. We can verify that both $E\sim \theta$ relation curve can be well fitted by the dashed lines described by the relations
\begin{equation}\label{energy_function}
E\left(\theta\right)=E_0+\eta\cos 2\theta
\end{equation}
with $\eta>0$. Consequently, the minimized energy is realized at $\theta=\pm\pi/2$, leading to $\psi_1:\psi_2=1:\pm i$. The Eq. (\ref{energy_function}) and the consequent $1:\pm i$ mixing manner have been understood by the G-L theory in the first section.

Therefore, the ground-state pairing gap functions of the two pairing symmetries take the form of $d_{x^2-y^2}\pm i d_{xy}$- or $g_{x^4+y^4-6x^2y^2}\pm i g_{x^3y-xy^3}$, abbreviated as $d+id$ and $g+ig$. Their amplitude and phase on the inner and outer Fermi surfaces when $\delta=0.32$ are shown in Fig.~\ref{mixing}. This figure shows that the obtained pairing states are fully-gapped without gap node. When the system is rotated by each $\pi/6$, the gap phases for the two pairing symmetries are shifted by $2\pi/6$ and $4\pi/6$ respectively. For each run around the FSs, the phase-distribution patterns for the two pairing symmetries repeat two or four times, leading to the winding numbers 2 and 4, respectively. Considering the presence of two Fermi pockets, the Chern numbers of the two pairing symmetries should be doubled\cite{6, 7}, given 4 and 8 for the $d+id$- and $g+ig$- wave pairings, respectively.

The topological properties of the obtained TSCs are robust against slight deviation of the twist angle from $30\degree$ to, say 29.9$\degree$. Under such deviation, the point group decays to $D_6$, and our solution of Eq. (\ref{linear_eq}) at, say the doping level $\delta=0.32$, yields a leading pairing symmetry belonging to the 2D IRRP of $D_6$ with $L=2$. The $1:\pm i$ mixing of the two obtained degenerate basis functions leads to a distribution of the gap phase angles on the FS very similar as that of the $g+ig$-wave pairing shown in Fig.~\ref{mixing}, yielding the same topological Chern numbers. The obtained state can be thought of as an approximate $g+ig$-wave TSC, as they have the same topological properties. The cases for other dopings are similar. Therefore, the pairing phase diagram for the case with the twist angle 29.9\degree is topologically the same as that for the case with the twist angle 30\degree.

\begin{figure}[htbp]
	\centering
	\includegraphics[width=0.7\textwidth]{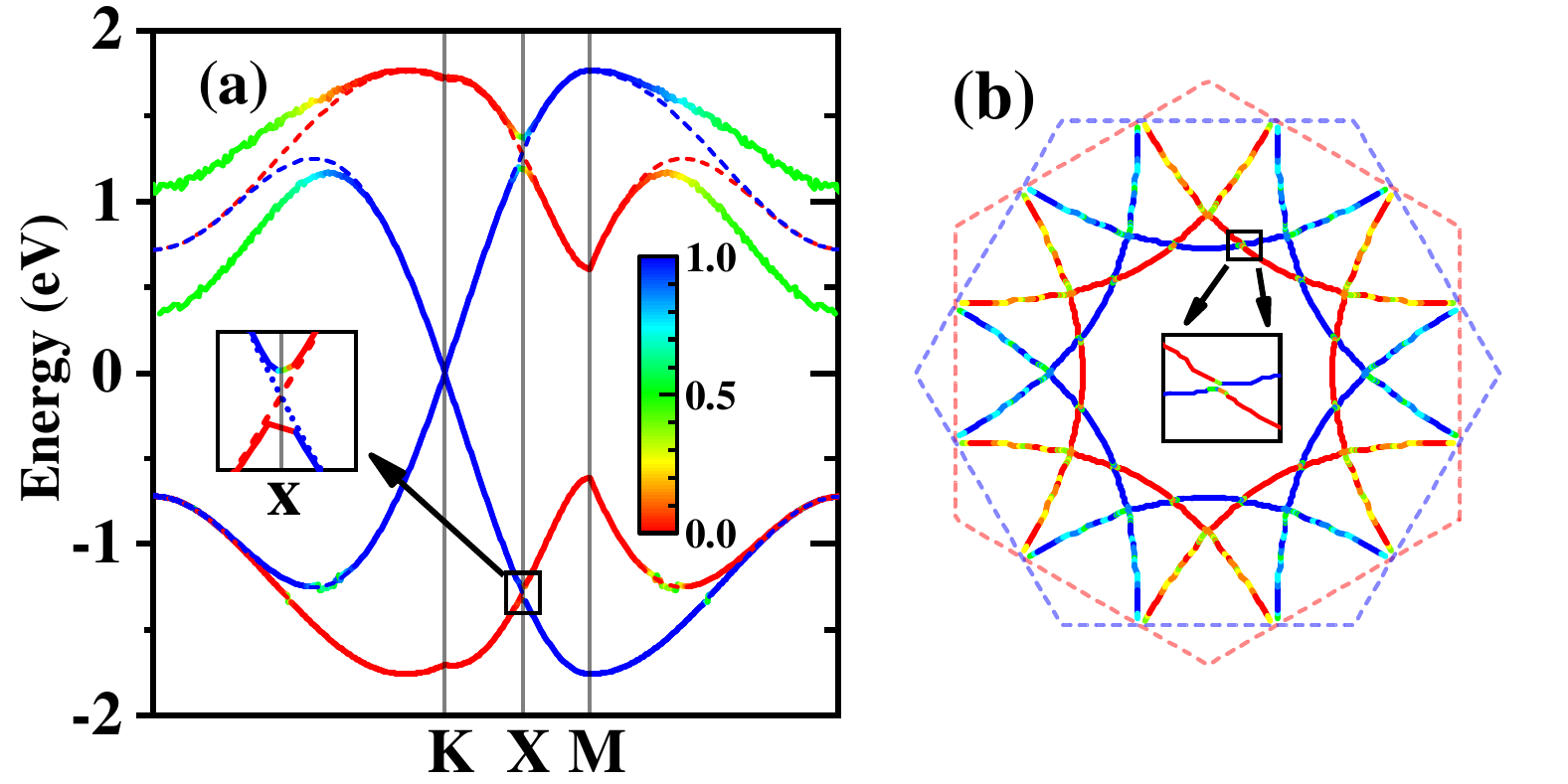}
	\caption{(a) Band structures of the 30\degree-twisted bilayer BC$_3$ with (solid lines) and without (dashed lines) the interlayer tunneling along the high-symmetry lines. (b) FSs of the 30\degree-twisted bilayer BC$_3$ for the doping level $\delta = 0.5$ electron doping in the Brillouin zone.}
	\label{energy-fs}
\end{figure}

\begin{figure}[htbp]
	\centering
	\includegraphics[width=0.65\textwidth]{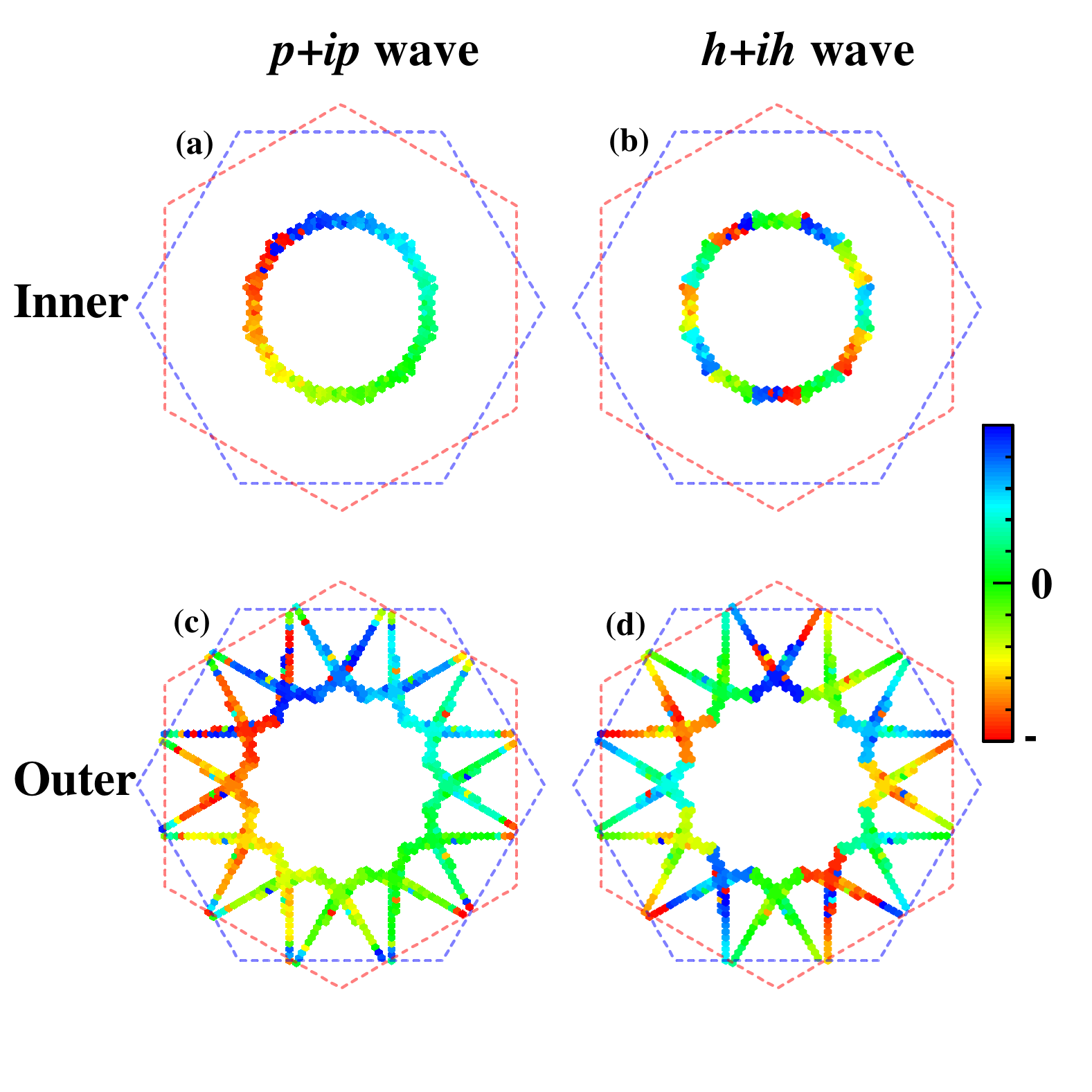}
	\caption{Distributions of the gap phases of the $p+ip$-wave and $h+ih$-wave pairings on the inner (a), (b) and outer (c), (d) FSs for $\delta=0.52$ electron doping in the 30\degree-twisted bilayer BC$_3$. }
	\label{phase}
\end{figure}

\subsection{The 30\degree-twisted bilayer BC$_3$}

For the 30\degree-twisted bilayer BC$_3$, the intralayer hopping integrals are provided by Ref\cite{bc3_thy}, including the nearest-neighbor (NN) hopping intergal $t_1=-0.62$ eV, the next-nearest-neighbor (NNN) one $t_2=0$ eV, and the next-next-nearest-neighbor (NNNN) one $t_3=0.38$ eV. The interlayer hopping integrals take the formula of Eq.~\eqref{tij}, with the related parameters given by $a\approx 0.297$ nm\cite{bc3_exp}, $d\approx 0.7$ nm, $t_{\pi}=-0.62$ eV, $t_\sigma=0.1$ eV, and $r_{0}=0.095$ nm, respectively. The interaction strength is given as $U=0.5$ eV\cite{bc3_thy}.

The Brillouin zone of the 30\degree-twisted bilayer BC$_3$ is the same as that in Fig.~\ref{dos-gap}(a). The corresponding band structure (solid lines) is in Fig.~\ref{energy-fs}(a), in comparison with the uncoupled band structures (dashed lines) from the two layers. The main effect of the interlayer coupling lies in that when the two uncoupled band branches cross the $\Gamma$-$X$ lines guaranteed by symmetry, see Fig.~\ref{energy-fs}(a), they would couple and hybridize, leading to the band splitting and the layer-component exchange. This is also reflected by the FS, see Fig.~\ref{energy-fs}(b). These behaviours are similar to the QC-TBG, comparing Fig.~\ref{energy-fs}(a) with Fig.~\ref{dos-gap}(b). The key difference between them is about the particle-hole (p-h) asymmetry: the strong interlayer coupling takes place in the electron-doped side for the QC-TBG but in the hole-doped side for the 30\degree-twisted bilayer BC$_3$. Such difference is caused by their different signs for the NN hopping integrals. This can be proved similarly as that in the subsection \ref{example-I}.

Figure~\ref{phase} shows the distributions of the gap phases of the obtained $p+ip$ and $h+ih$-wave pairings on the inner and outer FSs for $\delta=0.52$ near the VH doping. When the system is rotated by each $\pi/6$, the gap phases for the two pairing symmetries are shifted by $\pi/6$ and $5\pi/6$, leading to the winding number 1 and 5, respectively. Note that unlike the cases of $d+id$- or $g+ig$-wave pairings shown in Fig~\ref{mixing} where the distribution patterns repeat two or four times for each run around the FSs, here for the $h+ih$-wave pairing no exactly repeating pattern of the gap-phase distribution on the FSs is found, because the number 5 is not a divisor of 12 and thus the rotation by $2\pi/5$ is not a symmetry operation. However, the winding number 5 for the $h+ih$-wave pairing is still visible in Fig.~\ref{phase} (b) and (d).

\begin{figure}[htbp]
	\centering
	\includegraphics[width=0.6\textwidth]{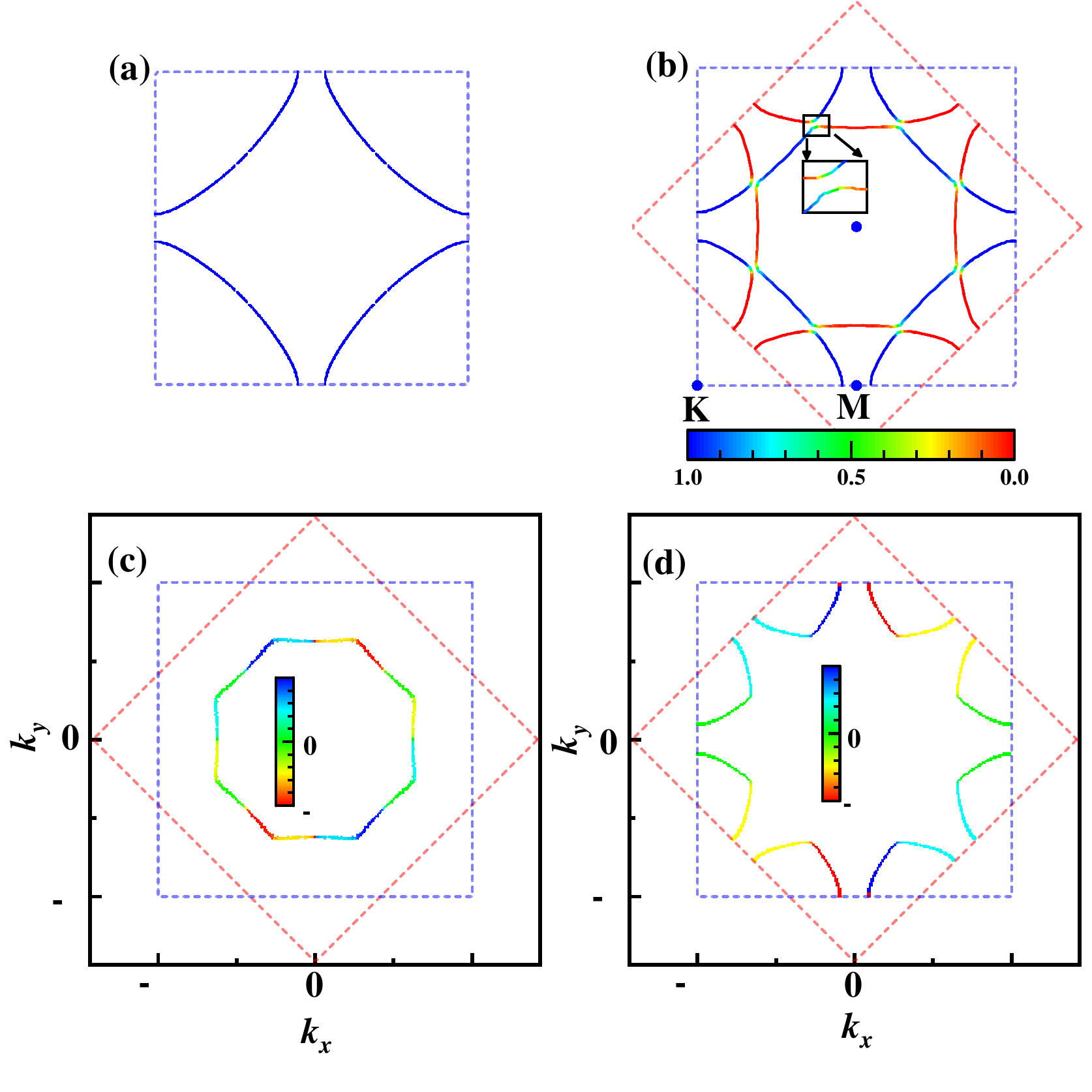}
	\caption{ FSs of (a) the single layer cuprates for the hole-doping $\delta=0.1$ and (b) the corresponding LA-THB. The distributions of the gap phase of the obtained $d+id$-wave pairing on the (c) inner and (d) outer FSs.}
	\label{cuprate}
\end{figure}

\subsection{The 45\degree-twisted bilayer cuprates superconductors}
For the 45\degree-twisted bilayer cuprates, the hopping integrals appearing in  Eq.~\eqref{tij} are set as follow. For the intralayer hopping integrals, we only keep the NN one set as $t_1=0.2$eV and the NNN one set as $t_2=-0.2t_1$. For the interlayer hopping integrals, we take the formula of Eq. (\ref{tij}), with the related parameters given by $a\approx0.54$ nm, $d\approx1.2$ nm, $t_{\pi}\approx0.68$ eV, $t_\sigma\approx30$ meV, and $r_{0}\approx0.211$ nm. The interlayer hopping integrals $\{t_{\mathbf{ij}}\}$ thus obtained are exactly the same as those given in Ref\cite{cuprate}. The superexchange interaction coefficients are set as $J_{\mathbf{ij}}=4t_{\mathbf{ij}}^2/U$, with $U=2$ eV to be the charge-transfer energy between the Cu-$3d$ orbitals and the O-$2p$ orbitals.

The FS of the single-layer system for the $\delta=0.1$ hole doping is shown in Figure~\ref{cuprate} (a), which only includes a hole pocket surrounding the $(\pi,\pi)$ point. The FSs for the corresponding QC LA-THB is shown in Figure~\ref{cuprate} (b), where the interlayer hybridization has made the FSs to split into inner and outer FSs. As clarified in the main text, our numerical results suggest that the degenerate $(d_{x^2-y^2}, d_{xy})$ doublets belonging to the 2D $E_2$ IRRP of the $D_8$ point group are the leading pairing symmetries, which are further mixed as $1:\pm i$ to lower the energy, forming the fully-gapped $d+id$ TSC. Figure~\ref{cuprate} (c) and (d) show the distribution of the gap phase of the obtained $d_{x^2-y^2}+i d_{xy}$-wave pairing on the inner and outer FSs, respectively. Since the two pockets are close in the Brillouin zone, the phase distributions of the gap functions are plotted on the inner and outer pockets separately to enhance the visibility. It can be clearly seen that the winding number is 2 in both the inner and outer pockets.

\end{widetext}

\begin{thebibliography}{10}
\bibitem{Kennes2021}
See, for example, the review in, D. M. Kennes, M. Claassen, L. Xian, A. Georges, A. J. Millis, J. Hone, C. R. Dean, D. N. Basov, A. N. Pasupathy and A. Rubio, Nat. Phys. 17, 155(2021).
\bibitem{Dean2018} R. Ribeiro-Palau, C. Zhang, K. Watanabe, T. Taniguchi, J. Hone, and C. R. Dean, Science 361, 690 (2018).
\bibitem{caoyuan20181}
Y. Cao, V. Fatemi, A. Demir, S. Fang, S. L. Tomarken, J. Y. Luo, J. D. Sanchez-Yamagishi, K. Watanabe, T. Taniguchi, E. Kaxiras, R. C. Ashoori, and P. Jarillo-Herrero, Nature 556, 80 (2018).
\bibitem{caoyuan20182}
Y. Cao, V. Fatemi, S. Fang, K. Watanabe, T. Taniguchi, E. Kaxiras, and P. Jarillo-Herrero, Nature 556, 43 (2018).
\bibitem{Chenguorui20191}
G.Chen, et al, Nature 572, 215(2019).
\bibitem{P_Kim2020}
X. Liu, Z. Hao, E. Khalaf, J. Y. Lee, K. Watanabe, T. Taniguchi, A. Vishwanath, and P. Kim, Nature 583, 221 (2020).
\bibitem{Chenshen2020}
C.Shen, et al, Nat. Phys. 16, 520(2020).
\bibitem{Park2021}
J. Park, et al, Nature 590, 249 (2021).
\bibitem{Xian2019}
L. Xian, D. M.Kennes, N. Tancogne-Dejean, M. Altarelli, and A. Rubio, Nano Lett. 19,4934(2019).
\bibitem{Wang2020}
L. Wang, et al, Nat. Mater. 19, 861(2020).
\bibitem{Regan2020}
E. C. Regan, et al,  Nature 579, 359(2020).
\bibitem{Tang2020}
Y. Tang, et al, Nature 579, 353(2020).
\bibitem{Yankowitz2019}
M. Yankowitz, S. Chen, H. Polshyn, Y. Zhang, K. Watanabe, T. Taniguchi, D. Graf, A. F. Young, and C. R. Dean, Science 363, 1059 (2019).
\bibitem{Yazdani2019}
Y. Xie, B. Lian, B. J{\"a}ck, X. Liu, C.-L. Chiu, K. Watanabe, T. Taniguchi, B. A. Bernevig, and A. Yazdani, Nature 572,101 (2019).
\bibitem{Efetov2019}
X. Lu, P. Stepanov, W. Yang, M. Xie, M. A. Aamir, I. Das, C. Urgell, K. Watanabe, T. Taniguchi, G. Zhang, A. Bachtold, A. H. MacDonald, and D. K. Efetov, Nature 574, 653 (2019).
\bibitem{David2019}
A. L. Sharpe, E. J. Fox, A. W. Barnard, J. Finney, K. Watanabe, T. Taniguchi, M. A. Kastner, D. Goldhaber-Gordon, Science 365, 605 (2019).
\bibitem{Serlin2019}
M. Serlin, C. L. Tschirhart, H. Polshyn, et al, Science 367,6480 (2019).
\bibitem{Zeldov2020}
A. Uri, S. Grover, Y. Cao, J. A. Crosse, K. Bagani, D. Rodan-Legrain, Y. Myasoedov, K. Watanabe, T. Taniguchi, P. Moon, M. Koshino, P. Jarillo-Herrero and E. Zeldov, Nature 581,47 (2020).
\bibitem{Caoyuan2021}
Y. Cao, D. Rodan-Legrain, J. M. Park, N. F. Yuan, K. Watanabe, T. Taniguchi, R. M. Fernandes, L. Fu, and P. Jarillo-Herrero, Science 372, 264 (2021).
\bibitem{Xu2018} C. Xu and L. Balents, Phys. Rev. Lett. \textbf{121}, 087001 (2018).
\bibitem{Po2018} H. C. Po, L. Zou, A. Vishwanath, and T. Senthil, Phys. Rev. X \textbf{8}, 031089 (2018).
\bibitem{Yuan2018} N. F. Q. Yuan and L. Fu, Phys. Rev. B \textbf{98}, 045103 (2018).
\bibitem{YangFan2018} C.-C. Liu, L.-D. Zhang, W.-Q. Chen, and F. Yang, Phys. Rev. Lett. \textbf{121}, 217001 (2018).
\bibitem{WuFeng20181} F. Wu, A. H. MacDonald, and I. Martin, Phys. Rev. Lett. \textbf{121}, 257001 (2018).
\bibitem{Kang2018} J. Kang and O. Vafek, Phys. Rev. X \textbf{8}, 031088 (2018); {\it ibid}, Phys. Rev. Lett. \textbf{122}, 246401 (2019).
\bibitem{Isobe2018} H. Isobe, N. F. Q. Yuan, and L. Fu, Phys. Rev. X \textbf{8}, 041041 (2018).
\bibitem{Koshino2018} M. Koshino, N. F. Q. Yuan, T. Koretsune, M. Ochi, K. Kuroki, and L. Fu, Phys. Rev. X \textbf{8}, 031087 (2018).
\bibitem{Fernandes2018} J. W. F. Venderbos and R. M. Fernandes, Phys. Rev. B \textbf{98}, 245103 (2018).
\bibitem{Dai2019} J. Liu, Z. Ma, J. Gao, and X. Dai, Phys. Rev. X \textbf{9}, 031021 (2019).
\bibitem{Gonzalez2019} J. Gonzalez and T. Stauber, Phys. Rev. Lett. \textbf{122}, 026801 (2019).
\bibitem{Song2019} Z. Song, Z. Wang, W. Shi, G. Li, C. Fang, and B. A. Bernevig, Phys. Rev. Lett. \textbf{123}, 036401 (2019).
\bibitem{Angeli2019} M. Angeli, E. Tosatti, and M. Fabrizio, Phys. Rev. X \textbf{9}, 041010 (2019).
\bibitem{Linyuping2019} Y.-P. Lin and R. M. Nandkishore, Phys. Rev. B \textbf{100}, 085136 (2019); {\it ibid}, Phys. Rev. B \textbf{102}, 245122 (2020).
\bibitem{MingXie2020} Ming Xie, A. H. MacDonald, Phys. Rev. Lett. \textbf{124}, 097601 (2020)
\bibitem{Abouelkomsan2020} A. Abouelkomsan, Z. Liu, and E. J. Bergholtz, Phys. Rev. Lett. \textbf{124}, 106803 (2020).
\bibitem{Bultinck2020prx} N. Bultinck, E. Khalaf, S. Liu, S. Chatterjee, A. Vishwanath, and M. P. Zaletel, Phys. Rev. X 10, 031034 (2020).
\bibitem{Senthil2020} C. Repellin, Z. Dong, Y.-H. Zhang, and T. Senthil, Phys. Rev. Lett. \textbf{124}, 187601 (2020).
\bibitem{Liao2021} Y. D. Liao, J. Kang, C. N. Breio, X. Y. Xu, H.-Q. Wu, B. M. Andersen, R. M. Fernandes, and Z. Y. Meng, Phys. Rev. X \textbf{11}, 011014 (2021).
\bibitem{Chichinadze2020} D. V. Chichinadze, L. Classen, and A. V. Chubukov, Phys. Rev. B \textbf{101}, 224513 (2020).
\bibitem{Mohammad2020} M. Alidoust, A.-P. Jauho, and J. Akola, Phys. Rev. Res. \textbf{2}, 032074(R) (2020).
\bibitem{Xian2021} L. Xian, et al, Nature Communications 12, 5644 (2021).
\bibitem{Angeli2020} M. Angeli and MacDonald, arxiv:2008.01735 (2020).
\bibitem{Naik2018} M. H. Naik, and M. Jain, Phys. Rev. Lett. 121, 266401 (2018).
\bibitem{Wu_TMD_2018} F. Wu, T. Lovorn, E. Tutuc and A. H. MacDonald, Phys. Rev. Lett. 121, 026402 (2018).
\bibitem{Ahn2018} S. J. Ahn, P. Moon, T.-H. Kim, H.-W. Kim, H.-C. Shin, E. H. Kim, H. W. Cha, S.-J. Kahng, P. Kim, M. Koshino, Y.-W. Son, C.-W. Yang, J. R. Ahn, Science \textbf{361}, 782 (2018).
\bibitem{Yao2018} W. Yao, E. Wang, C. Bao, Y. Zhang, K. Zhang, K. Bao, C. K. Chan, C. Chen, J. Avila, M. C. Asensio, J. Zhu, and S. Zhou, PNAS \textbf{115}, 6928 (2018).
\bibitem{Yan2019} C. Yan, D.-L. Ma, J.-B. Qiao, H.-Y. Zhong, L. Yang, S.-Y. Li, Z.-Q. Fu, Y. Zhang and L. He, 2D Mater. \textbf{6}, 045041 (2019).
\bibitem{Pezzini2020} S. Pezzini, V. Miseikis, G. Piccinini, S. Forti, S. Pace, R. Engelke, F. Rossella, K. Watanabe, T. Taniguchi, P. Kim and C. Coletti, Nano Lett. \textbf{20}, 3313 (2020).
\bibitem{Deng2020} B. Deng, B. Wang, N. Li, R. Li, Y. Wang, J. Tang, Q. Fu, Z. Tian, P. Gao, J. Xue and H. Peng, ACS Nano \textbf{14}, 1656 (2020).
\bibitem{Zhu2021} Yuying Zhu, Menghan Liao, Qinghua Zhang, Hong-Yi Xie, Fanqi Meng, Yaowu Liu, Zhonghua Bai, Shuaihua Ji, Jin Zhang, Kaili Jiang, Ruidan Zhong, John Schneeloch, Genda Gu, Lin Gu, Xucun Ma, Ding Zhang, and Qi-Kun Xue, Phys. Rev. X 11, 031011 (2021).
\bibitem{Zhao2021}
S. Y. Frank Zhao, N. Poccia, X. Cui, P. A. Volkov, H. Yoo, R. Engelke, Y. Ronen, R. Zhong, G. Gu, S. Plugge, T. Tummuru, M. Franz, J. H. Pixley, P. Kim, arXiv: 2108.13455.
\bibitem{Moon2013} P. Moon, M. Koshino, Phys. Rev. B \textbf{87}, 205404 (2013).
\bibitem{Koshino2015} M. Koshino, New J. Phys. \textbf{17}, 015014 (2015).
\bibitem{Moon2019} P. Moon, M. Koshino, and Y.-W. Son, Phys. Rev. B \textbf{99}, 165430 (2019).
\bibitem{Park2019} M. J. Park, H. S. Kim and S. B. Lee, Phys. Rev. B \textbf{99}, 245401(2019).
\bibitem{Crosse2021} J. A. Crosse and Pilkyung Moon, Phys. Rev. B \textbf{103}, 045408 (2021).
\bibitem{Yuanshengjun2019} G. Yu, Z. Wu, Z. Zhan, M. I. Katsnelson and S. Yuan, npj Comput Mater \textbf{5}, 122 (2019).
\bibitem{Yuanshengjun20201} G. Yu, M. I. Katsnelson and S. Yuan, Phys. Rev. B \textbf{102}, 045113 (2020).
\bibitem{Yuanshengjun20202} G. Yu, Z. Wu, Z. Zhan, M. I. Katsnelson and S. Yuan, Phys. Rev. B \textbf{102}, 115123 (2020).
\bibitem{Aragon2019} J. L. Aragon, G. G. Naumis and A. Gomez-Rodriguez, Crystals 9, 519 (2019).
\bibitem{SM} See the SM for the details of the pairing-symmetry classification, the G-L theory, the perturbational-band theory based microscopic framework and more informatios on the numerical results for the three exemplar systems.
\bibitem{exp} K. Kamiya, T. Takeuchi, N. Kabeya, N. Wada, T. Ishimasa, A. Ochiai, K. Deguchi, K. Imura, and N. K. Sato, Nat. Commun. \textbf{9}, 154 (2018).
\bibitem{DeGottardi2013} W. DeGottardi, D. Sen, and S. Vishveshwara, Phys. Rev. Lett. \textbf{110}, 146404 (2013).
\bibitem{YuPeng2013} X. Cai, L.-J. Lang, S. Chen, and Y. Wang, Phys. Rev. Lett. \textbf{110}, 176403 (2013).
\bibitem{Loring2016} I. C. Fulga, D. I. Pikulin, and T. A. Loring, Phys. Rev. Lett. \textbf{116}, 257002 (2016).
\bibitem{Sakai2017} S. Sakai, N. Takemori, A. Koga, and R. Arita, Phys. Rev. B \textbf{95},024509 (2017).
\bibitem{Andrade2019} R. N. Ara\'ujo and E. C. Andrade, Phys. Rev. B \textbf{100}, 014510 (2019).
\bibitem{Sakai2019} S. Sakai and R. Arita, Phys. Rev. Res. \textbf{1}, 022002(R) (2019).
\bibitem{Varjas2019} D. Varjas, A. Lau, K. Poyhonen, A. R. Akhmerov, D. I. Pikulin and I. C. Fulga, Phys. Rev. Lett. \textbf{123}, 196401 (2019).
\bibitem{Nagai2020} Y. Nagai, J. Phys. Soc. Jpn. 89, 074703 (2020).
\bibitem{Caoye2020} Y. Cao, Y. Zhang, Y.-B. Liu, C.-C. Liu, W.-Q. Chen, and F. Yang, Phys. Rev. Lett. \textbf{125}, 017002 (2020).
\bibitem{Sakai2020} N. Takemori, R. Arita, and S. Sakai, Phys. Rev. B \textbf{102}, 115108 (2020).
\bibitem{YYZhang2020} Y.-Y. Zhang, Y.-B. Liu, Y. Cao, W.-Q. Chen, and F. Yang, arXiv:2002.06485.
\bibitem{Hauck2021} J. B. Hauck, C. Honerkamp, S. Achilles, and D. M. Kennes, Phys. Rev. Res. \textbf{3}, 023180 (2021).
\bibitem{Zhoubin2021} C.-B. Hua, Z.-R. Liu, T. Peng, R. Chen, D.-H. Xu and B. Zhou, arXiv:2107.01439.
\bibitem{footnote2} If the pairing charalities from the two layers are opposite, the system cannot gain energy from the interlayer Josephson coupling. See the SM\cite{SM} for the argument.
\bibitem{MacDonald2011} R. Bistritzer and A. H. MacDonald, Proc. Natl. Acad. Sci. 108, 12233 (2011).
\bibitem{Castro2007} J. M. B. Lopes dos Santos, N. M. R. Peres and A. H. Castro Neto, Phys. Rev. Lett. \textbf{99}, 256802 (2007).
\bibitem{Doniach2007} A. M. Black-Schaffer, and S. Doniach, Phys. Rev. B \textbf{75}, 134512 (2007).
\bibitem{Gonzalez2008} J. Gonzalez, Phys. Rev. B \textbf{78}, 205431 (2008).
\bibitem{Honerkamp2008} C. Honerkamp, Phys. Rev. Lett. \textbf{100}, 146404 (2008).
\bibitem{Pathak2010} S. Pathak, V. B. Shenoy, and G. Baskaran, Phys. Rev. B \textbf{81}, 085431 (2010).
\bibitem{McChesney2010} J. L. McChesney, A. Bostwick, T. Ohta, T. Seyller, K. Horn, J. Gonzalez, and E. Rotenberg, Phys. Rev. Lett. \textbf{104}, 136803 (2010).
\bibitem{Nandkishore2012} R. Nandkishore, L. S. Levitov, and A. V. Chubukov, Nat. Phys. \textbf{8}, 158 (2012).
\bibitem{Wang2012} W.-S. Wang, Y.-Y. Xiang, Q.-H. Wang, F. Wang, F. Yang, and D.-H. Lee, Phys. Rev. B \textbf{85}, 035414 (2012).
\bibitem{Kiesel2012} M. L. Kiesel, C. Platt, W. Hanke, D. A. Abanin, R. Thomale, Phys. Rev. B \textbf{86}, 020507(R) (2012).
\bibitem{Honerkamp2014} A. M. Black-Schaffer and C. Honerkamp, J. Phys. Condens. Matter \textbf{26}, 423201 (2014).
\bibitem{exp_VHS} P. Rosenzweig, H. Karakachian, D. Marchenko, K. Kuster, and U. Starke, Phys. Rev. Lett. \textbf{125}, 176403 (2020).
\bibitem{BC3_exp} H. Yanagisawa, T. Tanaka, Y. Ishida, M. Matsue, E. Rokuta, S. Otani, and C. Oshima, Phys. Rev. Lett. 93, 177003 (2004).
\bibitem{BC3_doping} X. Chen and J. Ni, Phys. Rev. B 88, 115430 (2013).
\bibitem{BC3_VHS} Xi Chen, Yugui Yao, Hong Yao, Fan Yang, and Jun Ni, Phys. Rev. B 92, 174503 (2015).
\bibitem{VHS_II} H. Yao and F. Yang, Phys. Rev. B 92, 035132 (2015).
\bibitem{cuprates_QC} O. Can, T. Tummuru, R. P. Day, I. Elfimov, A. Damascelli, and M. Franz, Nat. Phys. \textbf{17}, 519(2021).
\bibitem{f_honeycomb} M. L. Kiesel, C. Platt, W. Hanke, D. A. Abanin, and R. Thomale, Phys. Rev. B \textbf{86}, 020507(R) (2012).
\end{thebibliography}

\begin{thebibliography}{99}

\bibitem{x3} W. Xu and X. Ka, "Group Theory and Application in Solid State Physics", Higher Education Press, Beijing (1999).
\bibitem{x2} R. Nandkishore, L. S. Levitov and A. V. Chubukov, Nat. Phys. \textbf{8}, 158 (2012).
\bibitem{1} P. Moon, M. Koshino and Y.-W. Son, Phys. Rev. B \textbf{99}, 165430 (2019).
\bibitem{2} W. Yao, E. Wang, C. Bao, Y. Zhang, K. Zhang, K. Bao, C.-K. Chan, C. Chen, J. Avila, M. C. Asensio, J. Zhu and S. Zhou, PNAS 115, 6928 (2018).
\bibitem{3} M. Koshino, New J. Phys. \textbf{17}, 015014 (2015).
\bibitem{Gutzwiller} F.-C. Zhang, C. Gros, T. M. Rice, and H. Shiba, Superconductor Science and Technology 1, 36 (1988).
\bibitem{4} S. Graser, T. A. Maier, P. J. Hirschfeld and D. J.Scalapino, New J. Phys. \textbf{11}, 025016 (2009).
\bibitem{5} T. A. Maier, S. Graser, P. J. Hirschfeld and D. J. Scalapino, Phys. Rev. B \textbf{83}, 100515(R) (2011).
\bibitem{x1} Y. Cao, Y. Zhang, Y.-B. Liu, C.-C. Liu, W.-Q. Chen and F. Yang, Phys. Rev. Lett. \textbf{125}, 017002 (2020).
\bibitem{graphene_RMP} A. H. Castro Neto, F. Guinea, N. M. R. Peres, K. S. Novoselov, and A. K. Geim, Rev. Mod. Phys. \textbf{81}, 109 (2009).
\bibitem{6} X.-L. Qi, T. L. Hughes and S.-C. Zhang, Phys.Rev.B \textbf{82}, 184516 (2010).
\bibitem{7} J. Alicea, Rep. Prog. Phys. \textbf{75}, 076501 (2012).
\bibitem{bc3_thy} X. Chen, Y. Yao, H. Yao, F. Yang, and J. Ni, Phys. Rev. B \textbf{92}, 174503 (2015).
\bibitem{bc3_exp} H. Yanagisawa, T. Tanaka, Y. Ishida, M. Matsue, E. Rokuta, S. Otani, and C. Oshima, Phys. Rev. Lett. \textbf{93}, 177003 (2004).
\bibitem{cuprate}  O. Can, T. Tummuru, R. P. Day, I. Elfimov, A. Damascelli, and M. Franz, Nat. Phys. \textbf{17}, 519(2021).

\end{thebibliography}
\end{document}